\documentclass[11pt, prd,nofootinbib,preprint,superscriptaddress]{revtex4}
\pdfoutput=1
\usepackage[T1]{fontenc}
\usepackage{amsmath,amssymb}
\usepackage{epsfig}
\usepackage{subfigure}
\usepackage{float}
\usepackage[normalem]{ulem}
\usepackage{graphicx}
\usepackage{slashed}
\usepackage{multirow}
\usepackage[colorlinks,citecolor=blue]{hyperref}
\usepackage{pdfpages}
\usepackage{color}
\usepackage{comment}
\usepackage[fleqn]{mathtools}

\definecolor{orange}{rgb}{1,0.5,0}

\DeclarePairedDelimiter\ket{\lvert}{\rangle}
\DeclarePairedDelimiterX\braket[2]{\langle}{\rangle}{#1\,\delimsize\vert\,\mathopen{}#2}

\begin{document}

\title{{\small Vector Dark Matter with Higgs Portal in Type II Seesaw framework}}

        \author{Nandini Das}
	\email{nandinidas.rs@gmail.com}
	\affiliation{School of Physical Sciences, Indian Association for the Cultivation of Science, 2A $\&$ 2B, Raja S.C. Mullick Road, Kolkata 700032, India}
 
	\author{Tapoja Jha}
	\email{tapoja.jha@oulu.fi}
         \affiliation{School of Physical Sciences, Indian Association for the Cultivation of Science, 2A $\&$ 2B, Raja S.C. Mullick Road, Kolkata 700032, India}
	\affiliation{Sodankylä Geophysical Observatory, University of Oulu, Tähteläntie, 99600 Sodankylä, Finland}
				
	\author{Dibyendu Nanda}
	\email{dibyendu.nanda@iopb.res.in}
	\affiliation{School of Physics, Korea Institute for Advanced Study, Seoul 02455, Republic of Korea}
   \affiliation{Institute of Physics, Sachivalaya Marg, Bhubaneswar 751005, India}

\begin{abstract}
We study the phenomenology of a vector dark matter (VDM) in a $U(1)_X$ gauged extension of the Standard Model (SM) which is set in a type II seesaw framework. When this $U(1)_X$ symmetry is spontaneously broken by the vacuum expectation value (VEV) of a newly introduced complex scalar singlet, the gauge boson $Z^\prime$ becomes massive. The stability of the dark matter (DM) is ensured by the presence of an exact charge conjugation symmetry resulting from the structure of the Lagrangian. On the other hand, the $SU(2)_L$ triplet scalar facilitates light neutrino masses via the type II seesaw mechanism. We have studied the phenomenology of the usual WIMP DM, considering all possible theoretical and experimental constraints that are applicable. Due to the presence of a triplet scalar, the present framework can also accommodate the observed $2\sigma$ deviation in $h \to Z \gamma $ decay rate, recently measured at the LHC. The possibility of non-thermal production of DM from the decay of the singlet scalar has also been briefly discussed.
\end{abstract}

\setcounter{footnote}{0}  
\renewcommand{\thefootnote}{\arabic{footnote}}  

\maketitle

\section{Introduction} \label{sec:intro}
The existence of a non-luminous, non-baryonic form of matter which provides roughly $26\%$ energy density of the universe, popularly known as the dark matter (DM), is an undeniable fact in terms of observational evidence. Astrophysical and cosmological observational evidence, notably the study of galaxy clusters by Fritz Zwicky \cite{Zwicky:1933gu}, the rotation curves of galaxy clusters by Vera Rubin  \cite{Rubin:1970zza} and the observation of bullet cluster by Chandra observatory \cite{Clowe:2006eq} have provided compelling evidence for the existence of this mysterious species in the
present universe. Additionally, gravitational lensing and anisotropy of cosmic microwave background radiation (CMBR) measured by several cosmological experiments \cite{WMAP:2006bqn,Roszkowski:2017nbc,Planck:2018vyg,Bertone:2004pz,Hu:2001bc} have further supported the need
for such enigmatic matter. The DM number density ($\Omega_{DM} h^2$) as reported by WMAP \cite{WMAP:2006bqn} and PLANCK \cite{Planck:2018vyg} is precisely measured to be $\Omega_{DM} h^2=0.12\pm 0.001$ at $68\%$ CL. However, the nature of DM remains a mystery. Though the observational evidence tells us about the gravitational interaction of the DM, the particle origin of it presents an intriguing possibility. If DM is an elementary particle, it is expected to exhibit some level of interaction with the Standard Model (SM). The natural approach would be to consider the weak interaction of DM particles with the SM bath. As an interesting fact, weakly interacting massive particles (WIMPs) of electroweak mass scale can be produced thermally in the early universe and then they can freeze out leaving the thermal relic density close to the DM abundance of the present day. This coincidence is often noted as WIMP miracle \cite{Kolb:1990vq}.

Other than the WIMP scenario, depending on the interaction strength, DM can be a FIMP (feebly interacting massive particle) \cite{Hall:2009bx} or a SIMP (Strongly interacting massive particle) \cite{PhysRevLett.113.171301,PhysRevD.41.2388} in nature. There are various other approaches as well namely super WIMP \cite{Feng:2003xh,Pospelov:2008jk}, secluded WIMP \cite{Pospelov:2007mp}, forbidden DM \cite{DAgnolo:2020mpt}. However, due to favourable detection possibilities at direct, indirect, and collider search experiments, WIMP is the most popular choice.
Direct detection (DD) experiments like LUX \cite{LUX:2016ggv}, XENONnT \cite{XENON:2018voc,XENON:2023cxc} and PANDAX \cite{PandaX-II:2016vec,PandaX-II:2017hlx} provides stringent bound on wide ranges of masses and couplings of WIMP. Recent bound from LZ experiment \cite{LZ:2022lsv} eliminates larger DM parameter space. 
WIMP scenario has been studied in various types of model set up. Depending on the choice of model, WIMP can be a Dirac or a Majorana fermion, a scalar (pseudo-scalar) or a vector boson. In comparison to fermions and scalars, vector gauge bosons as DM candidates have received a little lesser attention due to the necessity of adding extra gauge extensions. Vector dark matter (VDM) has been studied in \cite{Baek:2012se,Farzan:2012hh,Baouche:2021wwa,Nomura:2020zlm, Babu:2021hef,Barman:2017yzr} in various gauge extensions. Specifically, VDM in $U(1)$ extension  has been analysed in \cite{Duch:2015jta,Amiri:2022cbv}. In the above scenario, the dark matter couples to the SM through the Higgs portal via the mixing between the SM Higgs and the new scalar singlet. Here, we have discussed the variation of the VDM parameter space where the scalar sector has been extended with an additional triplet scalar. 
$SU(2)_L$ triplet Higgs seems to be a very obvious candidate in this context as the presence of a triplet in the scalar sector can facilitate neutrino mass thereby solving another important question of particle physics. The neutrino mass generation will be discussed in the next paragraph. The presence of the extra triplet will provide extra annihilation channels for the dark matter, thus it has more control in tuning of the relic abundance.

Neutrino mass is one of the most compelling reasons to go beyond the Standard Model (BSM).
Neutrino oscillation data from various experiments confirmed that neutrinos have tiny mass $\sim {\cal O} (0.1)$ eV  \cite{10.1093/ptep/ptaa104,Mohapatra:2005wg}. The most popular way to generate such 
tiny neutrino mass at tree level is the Seesaw mechanism. Based on the particle contents of the model, there are three types of Seesaw mechanisms. Type I seesaw \cite{Minkowski:1977sc,Yanagida:1979as,Gell-Mann:1979vob,Mohapatra:1979ia} extends the SM particle spectrum by three heavy right handed neutrinos where as type II \cite{MAGG198061, LAZARIDES1981287, PhysRevD.23.165, PhysRevD.22.2227, PhysRevD.22.2860,PhysRevLett.44.912,Ma:1998dn} adds a $SU(2)_L$ triplet scalar and type III \cite{Foot:1988aq,Ma:1998dx,Ma:2002pf,Hambye:2003rt} adds a $SU(2)_L$ triplet fermion. Type I seesaw models with an extra $U(1)_X$ symmetry have been studied in Ref.\,\cite{Okada:2018tgy,Okada:2019sbb,Okada:2020evk} in the context of dark matter phenomenology. 
Interestingly, in addition to the added advantage of vacuum stability, the type II seesaw mechanism can generate small neutrino masses with a $SU(2)_L$ triplet scalar with a very small vacuum expectation value (VEV) $v_t$ whereas the type I seesaw mechanism requires three right-handed Majorana neutrinos with masses in the range $10^9 -10^{10}$ GeV. Moreover, if we choose the triplet VEV $v_t$ to be in the range $1-3$ GeV( $> 10^{-4}$ GeV), we can get a mass range for the doubly charged and singly charged scalar that can be probed at colliders \cite{Ashanujjaman:2021txz}. Fermionic DM with type II seesaw in $U(1)_X$ model has been discussed in \cite{Okada:2022cby,Ghosh:2021khk}.

In this article, we explore the possibility of the existence of a gauge boson VDM in a general gauged $U(1)_X$ extension of SM where neutrino mass is generated by the type II seesaw mechanism. In addition to a triplet scalar, a $SU(2)_L$ singlet, the only particle that carries the $U(1)_X$ charge, is necessary to break this symmetry spontaneously resulting in a massive vector boson, $Z^\prime$ which plays the role of DM in this model. The stability of  DM is ensured by an exact charge conjugation symmetry which arises due to  $U(1)_X$ invariance. The interaction of the singlet scalar with the SM Higgs doublet acts as a portal between the dark sector and the visible sector. Moreover, the presence of a triplet scalar in this scenario gives rise to interesting DM phenomenology by mixing with the SM Higgs doublet and the scalar singlet and finally resulting into an enhanced DM parameter space. Therefore we study an interesting scenario that addresses the non-zero neutrino mass and DM abundance in a unified framework.  Another interesting factor that 
prompts us to explore this scenario is the recent combined report by 
the ATLAS and CMS collaboration on the $h \to Z \gamma$ decay channel. Their findings reveal 
a combined signal strength $\mu_{Z\gamma}  = 2.2 \pm 0.7 $ \cite{ATLAS:2023yqk}, indicating a slight 
$2\sigma $  discrepancy compared to the Standard Model prediction of 
$\mu_{Z\gamma}= 1 $. Some recent studies \cite{Barducci:2023zml,Boto:2023bpg} show a way to accommodate such excess by different model perspectives. Type II seesaw triplet (the singly charged and doubly charged component) can also accommodate non-standard values of $\mu_{Z\gamma}$ as shown in various studies \cite{Arbabifar:2012bd,BhupalDev:2013xol}. We show that in addition to the dominant contribution of the triplet charged scalar components, the new couplings introduced by the complex scalar singlet can affect the signal strengths $\mu_{\gamma \gamma}$ and $\mu_{Z\gamma}$.

 This article is organized as follows. In Sec.\ref{sec:model}, we describe the model in detail along with the generation of neutrino mass by the type II seesaw mechanism. In Sec.\,\ref{sec:ThCMP}, the experimental and theoretical constraints on the model parameters are discussed. In Sec.\,\ref{sec:dm}, we finally discuss the phenomenology of the thermal VDM  with the prediction of possible experimental signature of this allowed model parameter space from DD. In Sec.\,\ref{sec:freezein}, the possibility of freeze-in scenario of the VDM is discussed briefly. Finally, the conclusion is drawn in Sec.\,\ref{sec:summ}.
\section{Basic framework} \label{sec:model}
Let us now introduce the detailed framework of this model. We are exploring an extension of the SM that involves an additional triplet ($\Delta$) to the scalar sector to facilitate the generation of neutrino mass through the type-II seesaw mechanism via a lepton number non-conserving interaction. Furthermore, a complex scalar singlet ($S$) has been introduced which is charged under an additional abelian $U(1)_X$ gauge symmetry. When $S$ acquires VEV ($v_s$), it spontaneously breaks $U(1)_X$ symmetry and makes the gauge boson ($Z^\prime$) massive which serves as a potential DM candidate of the universe in this model. The total Lagrangian of the model is given by
\begin{equation} \label{eq:Ltot}
\mathcal{L}_{\rm Tot} = \mathcal{L}_{\rm SM} + \mathcal{L}_{\rm Scalar} + \mathcal{L}_{Z^{\prime}_{\rm Dark}}+\mathcal{L}^{\Delta}_{\rm Yuk},
\end{equation}
where, $\mathcal{L}_{\rm SM}$ represents the SM Lagrangian without the scalar sector.
The different BSM interaction terms are analysed in detail in the following sections.
\subsection{Scalar Sector} \label{subsec:PhyscalarF}

The most general renormalizable Lagrangian involving a triplet, a singlet scalar along with the SM Higgs doublet can be written as   
\begin{equation} \label{eq:Lscalar}
\mathcal{L}_{\rm Scalar} = {\mathcal{L}^{\rm Kin}_{\rm Scalar}} -V(\Phi,\Delta,S),
\end{equation}
where the kinetic part of scalar fields is given by
\begin{equation} \label{eq:LSKin}
{\mathcal{L}^{\rm Kin}_{\rm Scalar}} = (D^{ \mu}\Phi)^{\dagger}(D_{\mu}\Phi)\;+\;{\rm Tr}[(D^{ \mu}{\Delta})^{\dagger}(D_{\mu}{ \Delta})]\;+\;(D^{ \mu}S)^{*}(D_{\mu}S),
\end{equation}
where,
\begin{eqnarray} \label{eq:Dmu}
D_{\mu}\Phi & = & \partial_\mu\Phi\;+\;\frac{ig}{2}{W_{\mu}}^a \sigma^a \Phi \;+\; \frac{ig^\prime}{2}B_{\mu}\Phi, \\
D_{\mu}{ \Delta} & = & \partial_\mu{ \Delta}\;+\;\frac{ig}{2}[W_{\mu}^a \sigma^a,{ \Delta}]\;+\;ig^\prime B_{\mu}{ \Delta}, \\
D_{\mu}S & = & \partial_\mu S\;+\;i g_{x} \;{ x_s} Z_{\mu}^{\prime} S.
\end{eqnarray}
Here $g_x$ is the gauge coupling of $U(1)_X$ and $x_s$ is the $U(1)_X$ charge of complex scalar singlet $S$ which is taken to be unity for the whole analysis. The covariant derivative of $S$ takes an important role in the DM dynamics and has been discussed in the dark sector.  The expressions of the SM Higgs doublet $\Phi$, the scalar triplet $\Delta$ and the $S$ are given by, 
\begin{equation} \label{eq:scalar}
   \Phi = \begin{pmatrix}
    \phi^+ \\ \phi^0 
    \end{pmatrix},
   ~~
    { \Delta} = \begin{pmatrix}
   \frac{\Delta^+}{\sqrt{2}} & \Delta^{++}\\
    \Delta^{0} & -\frac{\Delta^+}{\sqrt{2}}
    \end{pmatrix},~~
     S = (S_{r} + i\; s_{i} ). 
\end{equation}
The scalar potential is given by
\begin{eqnarray}
  V(\Phi, { \Delta}, S) &=&  {\mu}_{\Phi}^2(\Phi^{\dagger}\Phi) + \lambda(\Phi^{\dagger}\Phi)^2 + M_{\Delta}^2 {\rm Tr}[{ \Delta}^{\dagger}{ \Delta}] + \lambda_1 (\Phi^{\dagger}\Phi){\rm  Tr}[{ \Delta}^{\dagger}{ \Delta}] \nonumber \\ 
 &+& \lambda_2 ({\rm Tr}[{ \Delta}^{\dagger}{ \Delta}])^2 +\; \lambda_3 {\rm Tr}[({ \Delta}^{\dagger}{ \Delta})^2] + \lambda_4(\Phi^{\dagger}{ \Delta}{ \Delta^{\dagger}}\Phi) + [\mu\; \Phi^{\rm T} i\sigma_2 { \Delta}^{\dagger}\Phi +\; {\rm h.c.}]
 \nonumber \\
 &+& {\mu}_{s}^{2} (S^{\dagger}S) + \lambda_s(S^{\dagger}S)^2 + \lambda_{s \phi}(S^{\dagger}S)({\Phi}^{\dagger}\Phi) + \lambda_{S\Delta}
(S^{\dagger}S){\rm Tr}[{ \Delta}^{\dagger}{ \Delta}].
\end{eqnarray} \label{eq:Spotn1}
 After electroweak symmetry breaking (EWSB), the scalars can be expanded around VEV as
\begin{equation} \label{eq:vevs}
    \phi^0 = \frac{1}{\sqrt{2}}(v_{d} + h_{0} + i \;g_{0}), ~~ \Delta_0 = \frac{1}{\sqrt{2}}(v_{t} + \delta_0 + i\;\eta_0), ~~ S = \frac{1}{\sqrt{2}}(v_{s} + s_r + i\;s_i).
\end{equation}
The scalar kinetic terms of Eqn.\,\ref{eq:LSKin} give rise to the masses of $W^{\pm}$, $Z_{\mu}$ and $Z_{\mu}^{\prime}$ after symmetry breaking as the following
\begin{equation}
\label{eqn:wz mass}
    m^2_W = \frac{1}{4} g^2 (v^2_d+2 v^2_t), ~~
    ~~ m^2_Z=\frac{1}{4} \frac{g^2}{cos^2\theta_W} (v^2_d+4 v^2_t), 
    ~~
    M^2_{Z^\prime}=g^2_x v^2_s.
\end{equation}
where, $\theta_{W}$ is the Weinberg angle.
The triplet VEV ($v_{t}$) gives an additional contribution to both $W^\pm$ mass and $Z$ boson mass and in turn contributes to $\rho$ parameter (discussed in Sec.\,\ref{sec:ThCMP}). 
The minimization of the scalar potential yields the following relations:
\begin{eqnarray} \label{eq:min}
\mu_{\Phi}^2 & = & - (v_{d}^{2}\lambda + \frac{v_{t}^{2}}{2}(\lambda_{1} + \lambda_{4}) + \frac{v_{s}^{2}}{2}\lambda_{s\phi} - \sqrt{2}v_{t}\mu), \\
M_{\Delta}^{2} & = & - (\frac{v_{d}^{2}}{2}(\lambda_{1} + \lambda_{4}) + v_{t}^{2}(\lambda_{2} + \lambda_{3}) + \frac{v_{s}^{2}}{2}\lambda_{S\Delta} - \frac{v_{d}^{2}\mu}{\sqrt{2}v_{t}}),\\
\mu_{s}^{2} & = & -\frac{1}{2}(v_{d}^{2}\lambda_{s \phi} + v_{t}^{2}\lambda_{S\Delta}) - v_{s}^{2}\lambda_{s}.
\end{eqnarray}
Using the above minimization conditions,
the mass-squared of the doubly-charged states $\Delta^{\pm\pm}$ is given by
\begin{equation} \label{eq:mDCH}
m_{{H}^{\pm\pm}}^{2} = \frac{v_{d}^{2}\mu}{\sqrt{2}v_{t}} - \frac{v_{d}^{2}\lambda_{4}}{2} - v_{t}^{2}\lambda_{3}.
\end{equation}
The charged scalar mass matrix in the basis of ($\phi^{\pm},~\Delta^{\pm}$) is noted as 
\begin{equation}
\mathcal{M}^2_C=\begin{pmatrix} - \frac{v_{t}^{2}\lambda_{4}}{2} + \sqrt{2}v_{t}\mu & \frac{v_{d}}{4}(\sqrt{2}v_{t}\lambda_{4} - 4\mu) \\ \frac{v_{d}}{4}(\sqrt{2}v_{t}\lambda_{4} - 4\mu) & -\frac{v_{d}^{2}}{4 v_{t}}(v_{t}\lambda_{4} - 2\sqrt{2}\mu) \end{pmatrix}
\end{equation}
The mass eigen states ($G^{\pm},~H^{\pm}$) in terms of the gauge eigen states can be written as 
\begin{equation} \label{eq:DiagSCH}
\begin{pmatrix} G^{\pm} \\ H^{\pm} \end{pmatrix} = \begin{pmatrix} \cos\theta & \sin\theta \\ -\sin\theta & \cos\theta \end{pmatrix} \begin{pmatrix} \phi^{\pm} \\ \Delta^{\pm} \end{pmatrix}
\end{equation}
where $\theta$ is the mixing angle and reads as
\begin{equation}
    \tan\theta = \frac{\sqrt{2} v_t}{v_d}
\end{equation}
The charged Goldstone gives mass to $W^{\pm}$. The mass eigenvalue of the physical scalar $H^{\pm}$ is as follows 
\begin{equation} \label{eq:mSCH}
 m^2_{H^{\pm}}=\frac{v_{d}^{2} + 2 v_{t}^{2}}{4 v_{t}}(2\sqrt{2}\mu - v_{t}\lambda_{4}) .
\end{equation}
The CP odd mass matrix in the basis of  ($g_{0},~\eta_{0}$) is given by
\begin{equation}
    \mathcal{M}^2_A= \begin{pmatrix}
          2\sqrt{2}v_{t}\mu &  -\sqrt{2}v_{d}\mu \\ -\sqrt{2}v_{d}\mu & \frac{v_{d}^{2}}{\sqrt{2}}\frac{\mu}{v_{t}}
    \end{pmatrix}
\end{equation}
where the off-diagonal entry arises from the lepton number violating term of the scalar potential and introduces mixing among the states. This matrix can be diagonalized by a rotation of the states by an orthogonal matrix. The mass eigenstates can be expressed as 
\begin{equation} \label{eq:DiagCPO}
\begin{pmatrix} \zeta \\ A \end{pmatrix} = \begin{pmatrix} \cos\beta & \sin\beta \\ -\sin\beta & \cos\beta \end{pmatrix} \begin{pmatrix} g_{0} \\ \eta_{0} \end{pmatrix},
\end{equation}
where $\beta$ is the mixing angle and is given by 
\begin{equation}
    \tan\beta =\frac{2 v_t}{v_d} 
\end{equation}
Here, $\zeta$ is the Goldstone mode which eventually makes the $Z$ boson massive. $A$ is the physical CP odd Higgs with mass eigen value,
\begin{equation} \label{eq:MA}
m_{A} = \sqrt{\frac{v_{d}^{2} + 4 v_{t}^{2}}{\sqrt{2} v_{t}}\mu}.
\end{equation}
 The CP even mass matrix in the basis of ($h_0$, $\delta_0$, $s_{r}$) is given by
\begin{equation}
    \mathcal{M}^2_{E}=\begin{pmatrix}
      2 v_{d}^{2}\lambda & v_{d} v_{t}\Big(-\frac{4 m_{H^{\pm}}^{2}}{v_{d}^2 + 2 v_{t}^{2}} + \frac{2 m_{A}^{2}}{v_{d}^2 + 4 v_{t}^{2}} + \lambda_{1} \Big)& v_{d} v_{s} \lambda_{s\phi}\\
     v_{d} v_{t}\Big(-\frac{4 m_{H^{\pm}}^{2}}{v_{d}^2 + 2 v_{t}^{2}} + \frac{2 m_{A}^{2}}{v_{d}^2 + 4 v_{t}^{2}} + \lambda_{1} \Big) &~~-2 m_{H^{\pm\pm}}^{2} + v_{d}^{2}\Big(\frac{4 m_{H^{\pm}}^{2}}{v_{d}^2 + 2 v_{t}^{2}} - \frac{m_{A}^{2}}{v_{d}^2 + 4 v_{t}^{2}} \Big) + 2v_{t}^{2}\lambda_{2}~~ & v_{s} v_{t} \lambda_{S\Delta} \\
     v_{d} v_{s} \lambda_{s\phi} & v_{s} v_{t} \lambda_{S\Delta} &  2 v_{s}^{2}\lambda_{s}
    \end{pmatrix}.
\end{equation}
 Gauge basis to mass basis transformation can be obtained by
\begin{equation} \label{eq:DiagNHS}
\begin{pmatrix} h \\ H_{1} \\ H_{2} \end{pmatrix} = \mathcal{O}_{\alpha} \begin{pmatrix} h_{0} \\ \delta_{0} \\ s_{r} \end{pmatrix},~~{\rm where}~~  \mathcal{O}_{\alpha} = R_{3}.R_{2}.R_{1},
\end{equation}
with,
\begin{equation}\hspace*{-0.3cm} \label{eq:R1R2R3}
R_{1} = \begin{pmatrix} 
\cos\alpha_{1} & \sin\alpha_{1} & 0 \\ -\sin\alpha_{1} & \cos\alpha_{1} & 0 \\ 0 & 0 & 1  \end{pmatrix},~~
R_{2} = \begin{pmatrix} \cos\alpha_{2} & 0 & \sin\alpha_{2} \\ 0 & 1 & 0 \\ -\sin\alpha_{2} & 0 & \cos\alpha_{2}  \end{pmatrix},~~ R_{3} = \begin{pmatrix} 1 & 0 & 0 \\  0 & \cos\alpha_{3} & \sin\alpha_{3} \\ 0 &   -\sin\alpha_{3} & \cos\alpha_{3}\end{pmatrix}.
\end{equation}
Here ($h$, $H_1$, $H_2$) are the mass eigen states. The gauge eigen states in terms of mass eigen states can be expressed as
\begin{eqnarray}
    h_0 &=& c_{\alpha_1} c_{\alpha_2} h - (c_{\alpha_3}  s_{\alpha_1} +
  c_{\alpha_1} s_{\alpha_2} s_{\alpha_3} ) H_1 + ( s_{\alpha_1} s_{\alpha_3}- 
  c_{\alpha_1} c_{\alpha_3} s_{\alpha_2}) H_2
  \label{a}\\
 \delta_0 &=& 
 c_{\alpha_2} s_{\alpha_1} h +(c_{\alpha_1} c_{\alpha_3} - 
 s_{\alpha_1} s_{\alpha_2} s_{\alpha_3}) H_1 - 
  (c_{\alpha_3} s_{\alpha_1} s_{\alpha_2} + c_{\alpha_1}  s_{\alpha_3} ) H_2 \\
s_r &=& 
  s_{\alpha_2} h + c_{\alpha_2} s_{\alpha_3} H_1+ c_{\alpha_2} c_{\alpha_3} H_2
\end{eqnarray}
In the above equations, $s_{\alpha_{i}}$ and $c_{\alpha_{i}}$ where $i = 1,~2,~3$, stand for $\sin\alpha_{i}$ and $\cos\alpha_{i}$ respectively. Scalar potential parameters ($\lambda$s) can be expressed in terms of the physical masses of the scalars, mixing angles, and VEVs. Such relations, listed in Appendix \ref{ss:lambda}, help us to parameterize the model in terms of physical quantities. So the set of free parameters of this model is \{$\rm{sin\alpha_1}$, $\rm{sin\alpha_2}$, $\rm{sin\alpha_3}$, $m_{H_1}$, $m_{H_2}$, $m_A$, $m_{H^{\pm\pm}}$, $m_{H^{\pm}}$, $v_s$\}. We will later see that the triplet scalar masses would be almost degenerate from theoretical constraints. So the set of free parameters are
\begin{equation*}
   \{ \sin\alpha_1, \sin\alpha_2, \sin\alpha_3, m_\Delta, m_{S},v_s\}
 \end{equation*}
 where $m_{\Delta}$ represents masses of the scalars $m_{H^{\pm \pm}},m_{H^{\pm}}, m_{H_1}, m_{A}$ which have large triplet component in them and $m_S$ is the mass of scalar $H_2$ which is dominantly singlet. 
\subsection{Dark Sector}
Now let us discuss the role of the additional gauge boson which plays the role of DM in our framework.
Apart from the singlet scalar $S$, no other fields couple to this extra gauge boson $Z^\prime$. When the singlet scalar $S$ acquires a VEV after the spontaneous breaking of $U(1)_X$, $Z^\prime$ becomes massive. The mass of the DM is given by
\begin{equation*}
    M_{Z^\prime}=g_x v_s
\end{equation*}
We can choose either $g_x$ or $v_s$ to be a free parameter. For the constraints coming from the scalar sector, we would be using $v_s$ whereas in the DM phenomenology, we will translate the limit of $v_s$ in $g_x$ vs $M_{Z^\prime}$ plane.

The $U(1)_X$ invariant Lagrangian enjoys an additional dark conjugation symmetry \cite{Farzan:2012hh,Ma:2017ucp} described as 

\begin{eqnarray}
    Z^{(A)}_2 &:& Z^\prime_\mu \to -Z^\prime_\mu,~~~~~S \to S^* \nonumber \\
    Z^{(B)}_2 &:& Z^\prime_\mu \to -Z^\prime_\mu,~~~~~S \to -S^* \nonumber 
\end{eqnarray}
If we consider this symmetry to be exact, the kinetic mixing between $Z^\prime$ and $B$ field ($U(1)_Y$ gauge boson of the SM) is prohibited consequently. If we relax the symmetry and allow a gauge kinetic mixing term proportional to $Z_{\mu \nu} B^{\mu \nu}$ along with the canonical kinetic term for $Z^\prime_{\mu}$ in $\mathcal{L}_{Z^\prime_{Dark}}$, the kinetic mixing angle needs to be very small to make the $Z^\prime$ a viable decaying DM candidate. Nevertheless, we disregard that scenario and consider the dark charge conjugation symmetry to be exact in this framework. All SM fields along with the extra scalar triplet do not carry any $U(1)_X$ charge, hence they do not couple to the extra gauge boson. This ensures the stability of this new heavy gauge boson $Z^\prime$, rendering it a viable DM candidate in this scenario.
\vspace{0.1mm}
\subsection{Yukawa Lagrangian and neutrino mass} \label{subsec:LYuk}
In the present framework, the SM Yukawa Lagrangian is augmented by a Seesaw mass term involving the Higgs triplet and two lepton doublets as follows,
\begin{equation} \label{eq:LYuk}
\mathcal{L}_{\rm Yuk}^{\Delta} = - (Y_{\Delta})_{jk} L_{j}^{T} C i\sigma_{2}{ \Delta}L_{k} + {\rm h.c.},
\end{equation}
where $Y_{\Delta}$ is the $3 \times 3$ Yukawa matrix, $C$ is the charge conjugation matrix and $L$ is the SM lepton doublet. The following Majorana mass for neutrinos can be 
generated from the above Yukawa term, 
\begin{equation}
    m_\nu= \sqrt{2} \, Y_\Delta v_t~.
    \label{eq:smallvt}
\end{equation}
Since $v_{t} \sim \mathcal{O}(GeV)$ in our analysis and the neutrino mass scale is in the eV range, therefore $Y_\Delta$ must be very small. For neutrino mass scale $m_\nu \sim 0.8$ eV and $v_t \sim 1$ GeV, $Y_{\Delta} \sim 5.66 \times 10^{-10}$. Apart from the kinematics, the value of $Y_\Delta $ controls the decay mode of $H^{\pm\pm}$. 
The partial decay width of $H^{\pm \pm}$ to like sign dileptons being proportional to ${\mid Y_\Delta \mid }^{2}$, $H^{\pm \pm} \to \ell^{\pm} \ell^{\pm}$ would be highly suppressed  for our choice 
of $v_t$ which corresponds to very small $Y_\Delta$ and  $H^{\pm \pm} \to W^\pm W^\pm$ channel will dominate consequently.
\section{Theoretical and experimental constraints on model parameters} \label{sec:ThCMP}
Now we are ready to discuss the possible experimental and theoretical constraints on the parameters of this model which would be used in our analysis.
\subsection{Experimental constraints}
\noindent $\bullet$ \textbf{ $\rho$ parameter constraint:} The triplet vacuum expectation value $v_t$ contributes to the $W^\pm $ and $Z$ boson masses at the tree-level (see Eqn.\,\ref{eqn:wz mass}), thus affecting the value of the $\rho$ parameter which is defined in terms of the VEVs $v_d$ and $v_t$ of the 
SM doublet and triplet scalars respectively as 
\begin{equation}
    \rho = \frac{m^2_W}{m^2_Z c^2_W}=\frac{1+\frac{2 v^2_t}{v^2_d}}{1+\frac{4 v^2_t}{v^2_d}}.
\end{equation}
The electroweak precision data constrains $\rho$ parameter to be very close to its SM value. From the latest data, it is constrained as $\rho =1.00038 \pm 0.00020$ \cite{10.1093/ptep/ptaa104} which signifies that at 3$\sigma$ level, $v_t$ should be $\lesssim$ 2.6 GeV.

\noindent $\bullet$ \textbf{Experimental limits on singly and doubly charged scalar masses:}
{This scenario predicts a plethora of heavy scalars that have been extensively searched 
at both $ e^+e^-$ and hadron colliders. The negative search results at LEP-2 puts a limit on the 
singly charged mass $m_{H^\pm} > 78 $ (GeV) \cite{LEPHiggsWorkingGroupforHiggsbosonsearches:2001ogs}. On the other hand, the lower limit on the doubly charged scalar mass $m_{H^{\pm\pm}}$ from the LHC is highly model dependent. 
The decay modes of the $H^{\pm \pm}$ depend on $v_t$ and the mass splitting $m_{H^{\pm\pm}} - m_{H^\pm}$. For $v_t < 10^{-4}$ GeV (large Yukawa coupling) and assuming degenerate scalars, $H^{\pm \pm }$ decays to like-signed dilepton with almost $100\%$ probability. The decay
mode of $H^{\pm\pm}$ changes rather drastically when $v_t > 10^{-4}$ GeV, leading to
various competing channels, like $H^{\pm} H^{\mp}$, $ W^\pm W^\mp $, and $W^\pm H^\mp$, if kinematically allowed. From current analysis at the LHC, $m_{H^{\pm \pm}}\gtrsim 1080$ GeV 
\cite{ATLAS:2022pbd} is allowed for $H^{\pm \pm} \to \ell^{\pm} \ell^{\pm}$ decay where the branching ratios to each of the possible leptonic final states ($ee,\mu \mu ,\tau \tau, e \mu, e \tau, \mu \tau$) are considered to be equal. However in a type II seesaw framework, when $H^{\pm \pm} \to W^{\pm} W^{\pm}$ decay is allowed, doubly charged Higgs mass has a lower bound of 420 GeV \cite{Ashanujjaman:2021txz}. In the following analysis, we have chosen $v_t \sim  \mathcal{O}$(1 GeV) and $m_{H^{\pm\pm}}$ to be greater than 420 GeV.}\label{xxx} 

{\noindent $\bullet$ \textbf{Collider constraints on the neutral scalar masses and mixing:}
The Higgs decay width measurement at LHC can constrain the mixing angle between the triplet and SM doublet. The upper bound on this mixing angle, $\sin\alpha_{1}$ is $\lesssim 0.05$ \cite{Bhattacharya:2017sml} which is consistent with the experimental observation of $h \to W W^*$ branching ratio. A similar bound is obtained on the mixing angle of the complex scalar singlet with SM Higgs from theoretical and experimental constraints. The most stringent bound comes from W mass correction at NLO \cite{Lopez-Val:2014jva}. For $m_S$ in the range $\sim \{200- 1000\}$ GeV, the upper bound on the mixing angle $\sin\alpha_{2}$ varies between ($0.2 - 0.3$) \cite{Robens:2016xkb}. In this model, the value of $\sin\alpha_{1}$ is taken to be $< 0.05$ and the maximum value of $\sin\alpha_{2}$ is taken to be $0.1$.} 
\label{constraint_1}

\noindent $\bullet$ \textbf{Constraint from Lepton Flavor Violation:}
The Higgs triplet in the model induces lepton flavor violating (LFV) decays at tree level. The branching ratio (BR) of the process $\ell_i \to \ell_j \ell_k \ell_m$ mediated by a doubly charged scalar  is given by
\begin{equation}
 {\rm BR} (\ell_i \to \Bar{\ell_j} \ell_k \ell_m)= \frac{|(Y_\Delta)_{ij} (Y_\Delta)_{km}|^2}{64 G^2_F m^4_{H^{\pm \pm}}}.
\end{equation}
Among all possible combinations from the above-mentioned branching ratios, the most stringent bound comes from $\mu \to e e e$ with BR($\mu \to e e e$) $\lesssim 1.0 \times10^{-12}$ from SINDRUM experiment \cite{SINDRUM:1987nra}. As Yukawa parameters are inversely proportional to $v_t$, we can extract a lower limit on $v_t$ from the above expression for a particular mass of $H^{\pm \pm}$. 
Another lepton flavor violating process, namely $\ell_i \to \ell_j \gamma $ receives non-standard contribution via loop diagrams of charged (singly and doubly) scalars. The BR of $\mu \to e \gamma$ gives the most stringent bound among these decays and the corresponding expression is given by \cite{Okada:2022cby}
\begin{equation}
   {\rm BR}(\mu \to e \gamma)= \frac{48 \pi^3 \alpha_{em}}{G^2_F m^2_{H^{\pm \pm}}}|({Y_\Delta}^\dagger Y_\Delta)_{e\mu} \frac{1}{16 \pi^2} \frac{3}{16}|^2.
\end{equation}
The latest bound on ${\rm BR}(\mu \to e \gamma) $ ($\lesssim 4.2 \times 10^{-13}$) has been recently reported by MEG collaboration \cite{MEG:2016leq}. In this model, $Y_\Delta$ being very small as mentioned before (see under Eqn.\,\ref{eq:smallvt}), the branching ratios of the aforementioned processes are way smaller than the reported experimental bounds.

\noindent $\bullet$ \textbf{Constraint from oblique parameters:}
Dominant contribution in oblique parameters comes from the charged Higgs sector of the triplet scalar. The most stringent bound comes from the $T$ parameter and it demands the mass splitting of the doubly charged and the singly charged scalars to be $\lesssim$ 40 GeV \cite{Chun:2012jw}. Our DM sector constraints are insensitive to this mass splitting, so the triplet scalar masses are taken to be degenerate for the analysis of the DM part.  

\noindent $\bullet$ \textbf{Higgs Invisible decay constraint:}
Current Higgs searches at the LHC puts stringent limits on the branching fraction of Higgs invisible decay. In our scenario, SM Higgs can decay to a pair of VDM through mixing with the singlet scalar when $m_h > 2 M_{Z^\prime}$. The expression of Higgs invisible branching ratio is given by
\begin{equation}
   {\rm BR}(h \to {\rm invisible}) = 
    \frac{\Gamma(h \to Z^\prime Z^\prime) }{\Gamma(h \to Z^\prime Z^\prime)+\Gamma(h \to {\rm SM~SM})}.
\end{equation}
This ratio is constrained to be $\le 10.5 \%$ by ATLAS collaboration \cite{ATLAS:2022yvh}. Here for $M_{Z^\prime}< 62.5$ GeV, the coupling of Higgs to DM will be constrained by this condition.

\noindent $\bullet$ \textbf{Constraints from $h \to \gamma \gamma $ and $h \to Z \gamma $ :}
In the SM, due to the absence of tree level coupling, $h \to \gamma \gamma $  takes place at one loop level via the exchange of charged 
fermions and $W^\pm $ bosons. This particular decay mode plays a significant role in the
discovery of the SM Higgs at the LHC. Being a loop-mediated process, it shows high sensitivity
to any new particles that couple to both the SM Higgs and the photon. 
Therefore, in the type II seesaw model, the existence of new heavy $H^{\pm}$ and $H^{\pm\pm}$ scalars has the potential to supplement the contributions to this particular channel. The impact of this additional contribution on the $\Gamma (h \to \gamma \gamma)$ can vary, either enhancing or reducing it, depending on the specific model parameters.
The partial decay width with these spin-0, spin-$1/2$, spin-1 contribution is given by \cite{Shifman:1979eb,Djouadi:2005gj,Spira:1997dg,Gunion:1989we,BhupalDev:2013xol}  
\begin{eqnarray}
     \Gamma(h\to \gamma \gamma)=\frac{\alpha^2 G_F m^3_h}{128 \sqrt{2} \pi^3}&&|\sum_f N_c Q^2_f g_{h f\Bar{f}} A^h_{1/2}(\tau_f)+ g_{h W^+ W^-} A^h_1 (\tau_W)+\Tilde{g}_{h H^\pm H^\mp} A^h_0 (\tau_{H^\pm}) \nonumber\\
     &+&4 \Tilde{g}_{h H^{\pm \pm} H^{\mp \mp}} A^h_0(\tau_{H^{\pm \pm}}) |^2 
\end{eqnarray}
where $\alpha$ is the fine structure constant, $G_F$ is the Fermi coupling constant, $N_c$ and $Q_f$ are the color factor and the electric charge of the fermion in the loop respectively. $\tau_i=\frac{m^2_h}{4 m^2_i}$ where $i$ corresponds to the virtual particles in the loop ($f,W,H^\pm, H^{\pm \pm}$ here). The expression of the couplings ($g_{h\,W^+\, W^-}$ etc.) and loop functions ($A^h_i(\tau_j)$ etc.) are given in the Appendix \ref{ss:htozgamma}. 

The first two terms on the right-hand side denote the SM fermions and W boson contribution in the loop whereas the last two terms represent the non-standard contribution coming from the singly charged ($H^\pm$) and the doubly charged ($H^{\pm \pm}$) scalars. Except for the coupling and mass dependence, the $H^{\pm \pm}$ term has a relative factor of four with respect to the $H^\pm$ term in the amplitude due to its enhanced charge.

Similarly, the partial decay width of $h \to Z \gamma$ is given by \cite{Arbabifar:2012bd,Carena:2012xa} 

\begin{eqnarray}
    \Gamma(h \to Z \gamma)&=&\frac{\alpha G^2_F m^2_W m^3_h}{64 \pi^2} (1-\frac{m^2_Z}{m^2_h})^3 \mid\frac{1}{c_W}\sum_f N_c Q_f(2 I^f_3- 4Q_f s^2_W) g_{h f\Bar{f}} A^h_{1/2}(\tau^f_h,\tau^f_Z) \nonumber \\
   && + c_W g_{h W^+ W^-} A^h_1 (\tau^W_h, \tau^W_Z) 
    +\frac{1}{s_W} g_{Z H^{\pm} H^{\mp}} \Tilde{g}_{h H^\pm H^\mp} A^h_0 (\tau^{H^\pm}_h, \tau^{H^\pm}_Z) \nonumber\\
    &&+ \frac{2}{s_W} g_{ZH^{\pm \pm} H^{\mp \mp}} \Tilde{g}_{h H^{\pm \pm} H^{\mp \mp}} A^h_0 (\tau^{H^{\pm \pm}}_h, \tau^{H^{\pm \pm}}_Z) \mid^2
\end{eqnarray}
where $\tau^i_j=\frac{4 m^2_i}{m^2_j}$ with $i=f,W,H^\pm,H^{\pm \pm}$ and $j=h,Z$. The necessary couplings and loop functions are mentioned in Appendix \ref{ss:htozgamma}.
\begin{table} [tbh!]
 \centering
   \begin{tabular}{c|c|c}
     \hline \hline
      Signal Strengths  & $\mu_{\gamma \gamma}$ & $\mu_{Z \gamma}$ \\
     \hline 
     ATLAS & $1.04^{+0.10}_{-0.09}$ \cite{ATLAS:2022tnm} & $2.0^{+1.0}_{-0.9}$ \cite{ATLAS:2020qcv} \\ 
      \hline
  CMS  &  $1.12^{+0.09}_{-0.09}$ \cite{CMS:2021kom} & $2.4^{-0.9}_{+0.9}$ \cite{CMS:2022ahq} \\
     \hline
   \end{tabular}
   \caption{Experimental signal strengths of $h \to \gamma \gamma$ and $h \to Z \gamma $ with uncertainties}
   \label{tab1}
  \end{table}
\begin{figure}[tbh!]
\centering
$$
\includegraphics[scale=0.35]{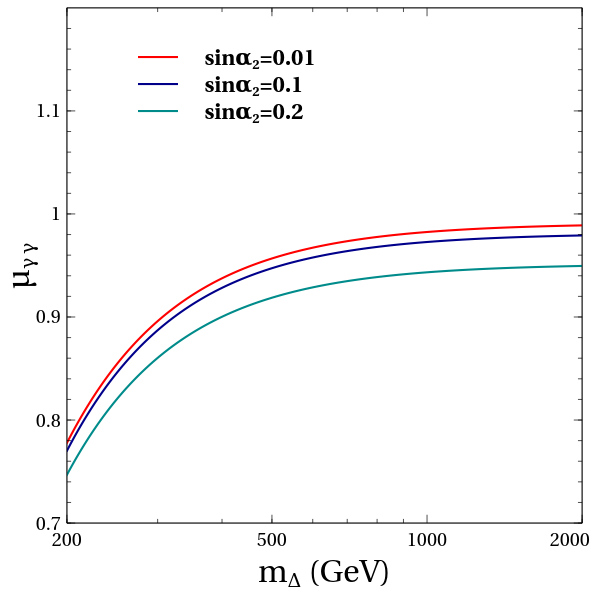}~~~~~~~~
\includegraphics[scale=0.35]{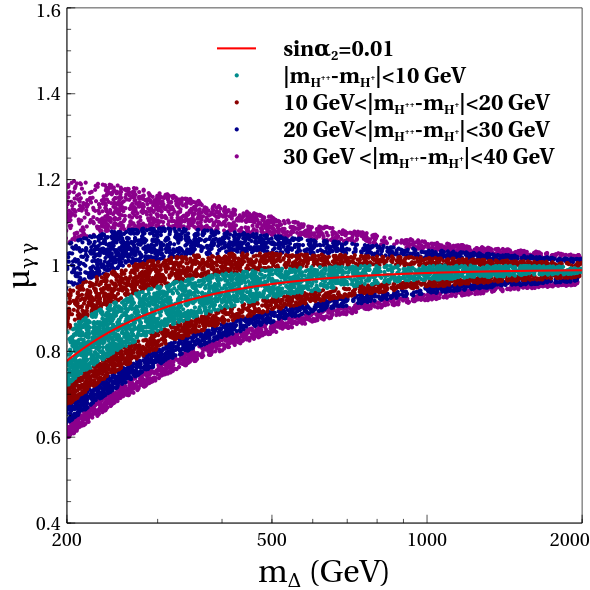}
$$
 \caption{ $h \to \gamma \gamma$ signal strength as a function of triplet scale $m_\Delta$. In the left panel, signal strength as a function of $m_\Delta$ ($=m_{H^{\pm \pm}},m_{H^\pm},m_{H_1},m_{A}$) for different values of $\sin\alpha_{2}$ mentioned inset.  In the right panel, the same plot for $\sin\alpha_{2}=0.01$ with different values of the mass difference between charged scalars is considered. The other parameters are fixed at $\sin\alpha_{1}$ is $0.008$, $\sin\alpha_{3}=0.001$, $v_s=500$ GeV, $m_S=200$ GeV taken for the plots.} 
 \label{DP}
\end{figure}

\begin{figure}
\centering
$$
\includegraphics[scale=0.35]{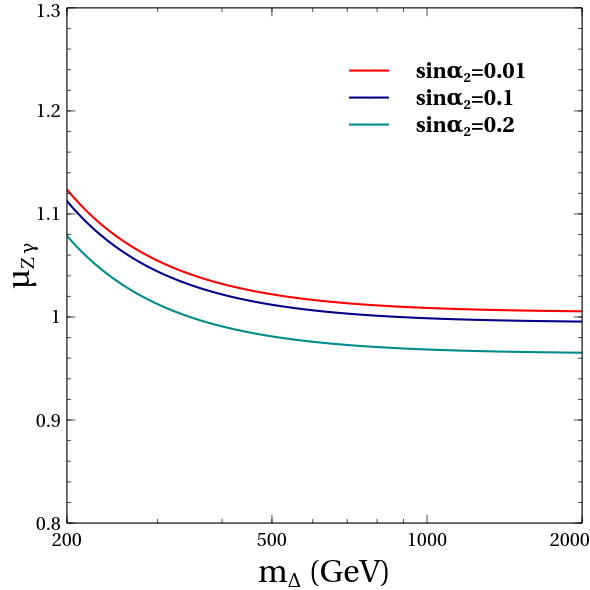}~~~~~~~~
\includegraphics[scale=0.35]{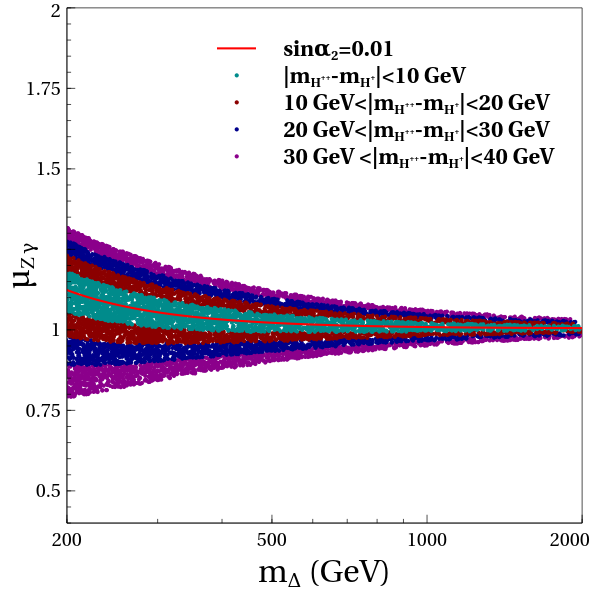}
$$
 \caption{ $h \to Z \gamma$ signal strength as a function of triplet scale $m_\Delta$. In the left panel, signal strength as a function of $m_\Delta$ ($=m_{H^{\pm \pm}},m_{H^\pm},m_{H_1},m_{A}$) for different values of $\sin\alpha_{2}$ mentioned inset.  In the right panel, the same plot for $\sin\alpha_{2}=0.01$ with different values of the mass difference between charged scalars is considered. The value of $\sin\alpha_{1}$ is $0.008$, $\sin\alpha_{3}=0.001$, $v_s=500$ GeV, $m_S=200$ GeV taken for the plots.}
 \label{ZP}
\end{figure}
\noindent The partial decay width of $h \to \gamma \gamma$ and $h \to Z \gamma$ in the type II seesaw extended model differs from SM due to the presence of a triplet scalar. In SM, the $W$ loop interferes with the top loop destructively. In the presence of triplet scalars, the doubly charged and singly charged scalars interfere constructively with the $W$ contribution, as a result, the decay width enhances. The effect of triplet scalar in $h \to  \gamma \gamma$ and $h \to Z \gamma$ signal has been discussed in Ref.\,\cite{Arbabifar:2012bd,BhupalDev:2013xol}. Here as another CP even scalar from the complex scalar singlet mixes with the Higgs, the coupling of Higgs to the charged scalars is modified with $\lambda_{\Delta S}$, $\lambda_{s\phi}$ along with mixing angles.
The signal strength of $h \to \gamma \gamma$ at LHC is given by
\begin{eqnarray}
    \mu_{\gamma \gamma}=\frac{\sigma_{\rm model}(pp \to h \to \gamma \gamma)}{\sigma_{\rm SM}(pp \to h \to \gamma \gamma}&=& \frac{\Gamma_{\rm model}(h \to gg)}{\Gamma_{\rm SM}(h \to gg)} \frac{{\rm Br}(h \to \gamma \gamma)_{\rm model}}{{\rm Br}(h \to \gamma \gamma)_{\rm SM}} \nonumber \\
 &=&   \cos\alpha^2_{1} \cos\alpha^2_{2} \frac{{\rm Br}(h \to \gamma \gamma)_{\rm model}}{{\rm Br}(h \to \gamma \gamma)_{\rm SM}} 
\end{eqnarray}
Similarly, we can write the signal strength of $h \to Z \gamma$ as
\begin{equation}
   \mu_{Z \gamma} =  \cos\alpha^2_{1} \cos\alpha^2_{2} \frac{{\rm Br}(h \to Z \gamma)_{\rm model}}{{\rm Br}(h \to Z \gamma)_{\rm SM}}
\end{equation}
The experimental bounds on the signal strengths are mentioned in Table.\ref{tab1}.
In Fig.\,\ref{DP}, the signal strength for $h \to \gamma \gamma$ is shown as a function of the triplet mass scale $m_\Delta$. In the left panel, $\mu_{\gamma \gamma}$ is plotted as a function of $m_\Delta$ for different values of $\sin\alpha_{2}$. In the right panel, the signal strength of $h \to \gamma \gamma$ is plotted as a function of $m_\Delta$ for $\sin\alpha_{2}=0.01$ when different mass gaps between the charged scalars are considered. Here we see that as the dominant couplings that contribute to the decay widths heavily depend on the charged scalar masses, consideration of the non-degenerate mass of the triplet scalars provides us with an enlarged parameter space which is consistent with the measurement of signal strength by ATLAS and CMS. Another important point to note is that consideration of the singlet scalar introduces an important parameter $\sin\alpha_{2}$ in the signal strength expression which can affect the value of signal strength $\mu_{\gamma \gamma}$ as shown in the right panel of Fig.\,1. Similar treatment for $\mu_{Z \gamma}$ is done in Fig.\,\ref{ZP}. Here we see that while non-degeneracy of the charged scalars is considered, the model parameter space for $m_\Delta$ up to  $600$ GeV predicts a signal strength that is within $1 \sigma$ deviation of the ATLAS measurement of signal strength and in $2 \sigma$ limit of the CMS measurement of signal strength. Our model parameter space lies within $2 \sigma$ limit of a recent combined analysis of ATLAS and CMS search data which show  $\mu_{Z \gamma}=2.2\pm 0.7$ at the center of mass energy $13$ TeV \cite{ATLAS:2023yqk}. 
\subsection{Theoretical constraints}
\noindent $\bullet$ \textbf{Vacuum stability conditions:} \label{subsec:bfb}
The stability of the potential demands that the quartic couplings and some of their combinations must be positive. 
 These conditions ensure that the potential is bounded from below in any direction. The necessary co-positivity conditions imposed on the model parameters are as follows
\begin{eqnarray} \label{eq:2field}
\lambda \geq 0,~~\lambda_{s} \geq 0,~~ \lambda_{2} + \lambda_{3} \geq 0, ~~2\lambda_{2} + \lambda_{3} \geq 0,\, \lambda_{s\phi} + 2\sqrt{\lambda \lambda_{s}} &\geq& 0,\\ 
\lambda_{S\Delta} + 2\sqrt{(\lambda_{2} + \lambda_{3})\lambda_{s}} \geq 0,\, \lambda_{S\Delta} + \sqrt{2(2\lambda_{2} + \lambda_{3})\lambda_{s}} &\geq& 0,\\
 (\lambda_{1} + \lambda_{4}) + 2\sqrt{(\lambda_{2} + \lambda_{3})\lambda} \geq 0,\, \lambda_{1} + 2\sqrt{\lambda(\lambda_{2} + \lambda_{3})} &\geq& 0,\\
 (2\lambda_{1} + \lambda_{4}) + 2\sqrt{2(2\lambda_{2} + \lambda_{3})\lambda} \geq 0,\,
\sqrt{2}(\lambda_{2} + \lambda_{3}) + \sqrt{(\lambda_{2} + \lambda_{3})(2\lambda_{2} + \lambda_{3})} &\geq& 0.
\end{eqnarray}

\noindent $\bullet$ \textbf{Perturbative unitarity:} \label{sebsec:unitarity}
It is also important to check wheather the $2\to 2$ scattering amplitudes involving the scalar fields satisfy unitarity constraints. Consequently the parameters of the scalar potential are constrained from above as the following \cite {PhysRevD.16.1519,PhysRevD.77.095009,GAKEROYD2000119}
\begin{eqnarray} 
 \Big\{|\lambda|,~ |\lambda_{2}|,~|\lambda_{s}|,~ |(\lambda_{2} + \lambda_{3})|\Big\}  \leq  4\pi,\, \Big\{|(2\lambda_{1} - \lambda_{4})|, |(2\lambda_{1} + 3\lambda_{4})|\Big\} &\leq& 16\pi, \nonumber\\
   \Big\{|\lambda_{1}|,~ |(\lambda_{1} + \lambda_{4})|,~ |(2\lambda_{2} - \lambda_{3})|,~ |\lambda_{s\phi}|,~ |\lambda_{S\Delta}|,\, |(\lambda + \lambda_{2} + 2\lambda_{3}) \pm \sqrt{(-\lambda + \lambda_{2} + 2\lambda_{3})^{2} + \lambda_{4}^{2}}|\Big\} &\leq& 8\pi \nonumber \\
\label{eq:unitarity}
\end{eqnarray}
The results noted in Eqn.\,\ref{eq:unitarity} have been obtained by a coupled channel analysis. However, some of the eigen values can not be written in closed form. For such cases, we evaluate the eigen values of the coupled channel matrix numerically and impose the unitarity constraints. The form of the matrix is noted in Appendix \ref{appendix:unitarity}.

   \subsection{Effect of constraints on model parameter space} \label{sec:theoretical constraints}
 The experimental and theoretical conditions described in Sec.\,\ref{sec:ThCMP} put stringent constraints on the model parameter space.
  The aforementioned constraints on quartic couplings, in turn, can be translated on the physical parameters like masses and mixing, and in the following, we present such model parameter space.
  
\begin{figure}[tbh]
      \centering
      \includegraphics[scale=0.35]{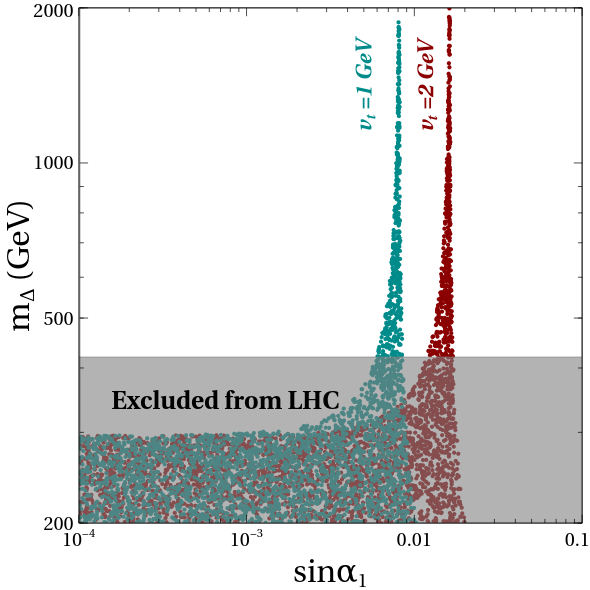}
      \caption{ Parameter space allowed from unitarity and vacuum stability conditions in $m_\Delta$ vs $\sin\alpha_{1}$ plane. The points with dark cyan colour (dark red) correspond to $v_t=1$ ($v_t=2$)GeV. Here  $m_S=200$ GeV and $v_s=1000$ GeV while the value of $\sin\alpha_{2}$ and $\sin\alpha_{3}$ are taken to be zero.}
      \label{fig:m_T}
  \end{figure}

The conditions on $\lambda$ parameters prefer near mass degeneracy among the triplet dominated scalars.  
 Fig.\,\ref{fig:m_T} shows the allowed parameter space from unitarity and vacuum stability in the $m_{\Delta}$ vs $\sin\alpha_{1}$ plane. Here we consider $\sin\alpha_{2}$ and $\sin\alpha_{3}$ to be zero implying that $S$ is completely decoupled from the triplet and the SM doublet. The magenta (cyan) coloured points correspond to the case when $v_t = 2$ ($v_t = 1$) GeV. The choice of $v_t$ has been made keeping it consistent with the $\rho$ parameter constraint. We see in this plot that for a specific value of $\sin\alpha_{1}$ corresponding to each value of $v_{t}$, all values of $m_{\Delta}$ are allowed where $\sin\alpha_{1} \sim (2 v_{t}/v_{d})$. For other values of $\sin\alpha_{1}$, $m_{\Delta} \leq 300$ GeV. This feature also has been observed in the Ref.\,\cite{Ghosh:2017pxl}. As $m_{H_1} \leq 300$ GeV is already excluded for the degenerate triplet dominated scalars from charged Higgs searches (see Sec.\,\ref{xxx}) (shown by the shaded region) and we are interested in the whole mass range of the triplet scalar for the analysis of DM phenomenology, we set $\sin\alpha_{1}$ as 0.008 for $v_{t}=1~{\rm GeV}$ for the rest of this article.

  \begin{figure}[tbh]
      \centering
      $$
      \includegraphics[scale=0.35]{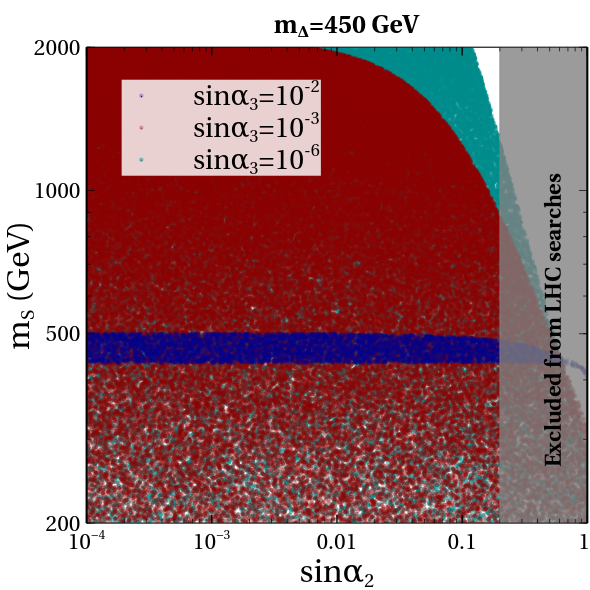}~~~~~~~~~~~~
      \includegraphics[scale=0.35]{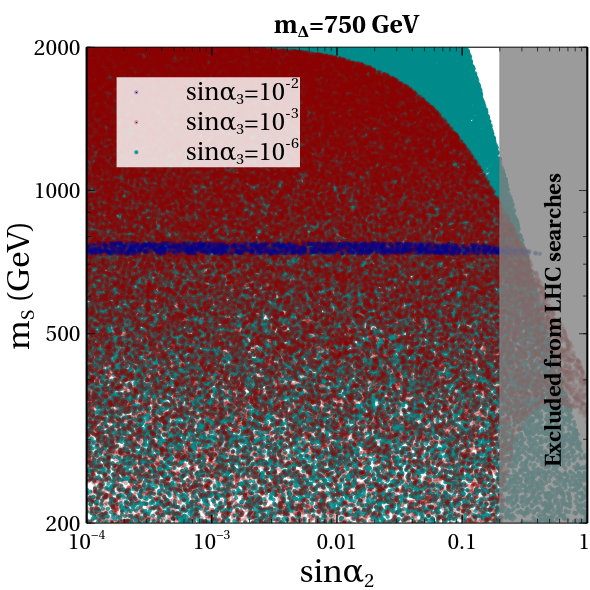}
      $$
      \caption{ Parameter space allowed from unitarity and vacuum stability conditions in $m_S$ vs $\sin\alpha_{2}$ plane. The different colours correspond to different values of $\sin\alpha_{3}$ (mentioned inset). The value of $v_s$ taken here is $1$ TeV. For the left panel, $m_\Delta=450$ GeV whereas for the right panel  $m_\Delta=750$ GeV. $\sin\alpha_1$ is set to be $0.008$ corresponding to $v_t=1$ GeV. }
      \label{fig:m3_s13}
  \end{figure}
  In Fig.\,\ref{fig:m3_s13}, we analyse the parametric dependence of $\sin\alpha_{3}$ and $m_{\Delta}$ on $m_{S}$ vs $\sin\alpha_{2}$ plane coming from unitarity and vacuum stability constraints. In the left panel (right), $m_{\Delta}$ has been fixed at 450 GeV (750 GeV). The different colours in the figure correspond to different values of $\sin\alpha_{3}$. It is evident that with higher values of mixing angle $\sin\alpha_{3}$, the allowed masses of $H_2$ ($m_S$) become close to the mass scale of $H_1$ ($m_\Delta$). As for example, in the left (right) panel, the allowed masses of $m_S$ are close to $450$ ($750$) GeV which is the scale of the triplet mass when $\sin\alpha_{3}=0.01$. 
  
\begin{figure}
      \centering
      \includegraphics[scale=0.4]{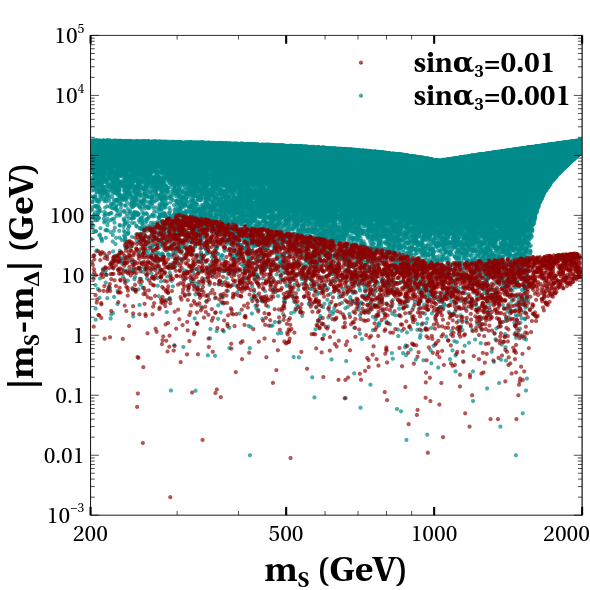}
      \caption{ Parameter space allowed from unitarity and vacuum stability conditions in $|m_S-m_{\Delta}|$ vs $m_S$ plane. The points with dark cyan (dark red) colour correspond to $\sin\alpha_{3}=0.001$ ($\sin\alpha_{3}=0.01$). The value of $m_\Delta$ taken here is $200$ GeV and $v_s=1000$ GeV while the value of $\sin\alpha_{1}$ and $\sin\alpha_{3}$ are taken to be $0.008$ and $0$ respectively. }
      \label{fig:m_S_m_T}
  \end{figure}

  In Fig.\,\ref{fig:m_S_m_T}, we observe the effect of mixing angle $\sin\alpha_{3}$ in $|m_{S}-m_{\Delta}|$ vs $m_{S}$ plane under the same theoretical constraints. Points with dark cyan colour (dark red) correspond to $\sin\alpha_{3}=0.001$ ($\sin\alpha_{3}=0.01$). We see that the unitarity and vacuum stability constraints allow a mass difference $|m_{S}-m_{\Delta}|\lesssim 100$ GeV when $\sin\alpha_{3}=0.01$ whereas for $\sin\alpha_{3}=0.001$, the allowed mass difference is $|m_{S}-m_{\Delta}|\lesssim 1000$ GeV. 
  
  Now we will study the effect of $v_s$ on our model parameter space.
  \begin{figure}
      \centering
      \includegraphics[scale=0.4]{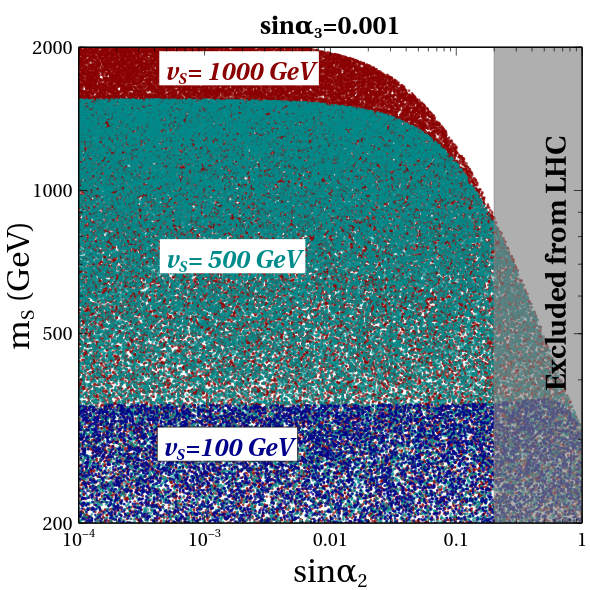}
      \caption{ Parameter space allowed from unitarity and vacuum stability conditions in $m_S$ vs $\sin\alpha_{2}$ plane. The points with different colours correspond to different values of $v_s$. The value of $m_\Delta$ taken here is $430 GeV$ and $\sin\alpha_{3}=0.001$. }
      \label{fig:vs-variation}
  \end{figure}
   In Fig.\,\ref{fig:vs-variation}, allowed parameter space from theoretical constraints is shown in $m_S$ vs $\sin\alpha_{2}$ plane. The allowed region in $m_S$ diminishes with smaller values of $v_s$ for all possible values of mixing angle, $\sin\alpha_{2}$. $\lambda_S$, $\lambda_{s\phi}$ and $\lambda_{S\Delta}$ become smaller as $v_s$ increases (see Eqn.\,\ref{ss:lambda}). Consequently,
   the unitarity constraints on such quartic parameters are automatically satisfied for higher values of $v_s$. This has been reflected in larger allowed regions in $m_S$ vs $\sin\alpha_{2}$ plane for larger values of $v_s$. The shaded region shown in Fig.\,\ref{fig:m3_s13} and Fig.\,\ref{fig:vs-variation} comes from the exclusion limit of the mixing angle of a scalar singlet with SM Higgs from the LHC as mentioned in Sec.\,\ref{constraint_1}.

\section{DM phenomenology} \label{sec:dm}
\begin{figure}[tbh]
\unitlength = 1mm
\hspace{3em}
$$
\includegraphics[scale=0.35]{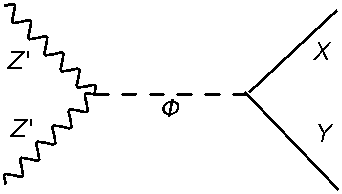}~~~~~~
\includegraphics[scale=0.35]{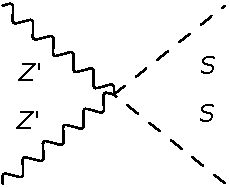}~~~~~~
\includegraphics[scale=0.35]{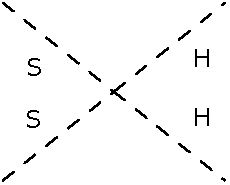}
$$
\centering
\medskip
\caption{ Feynman diagrams that contribute to the thermalisation of DM $Z^\prime$. X and Y represent SM particles along with $S$ ($H_2$) and triplet dominated scalars ($H_1,A_0,H^{\pm \pm},H^{\pm}$). $\phi$ denotes $h,H_1,H_2$.}
\label{xy}
\end{figure}

Now we are equipped to discuss the DM phenomenology in this model.  As a consequence of dark charge conjugation symmetry, $Z^\prime$ becomes stable and plays the role of DM in this framework. For a sizable value of new gauge coupling $g_x$, $Z^\prime$ can be present in the thermal bath of the early universe via its interaction (mediated by $S$) with the SM fields. However, this is possible only when $S$ itself is in thermal equilibrium with the SM fields. Thermalisation of $S$ is dominantly controlled by $\lambda_{s\phi}$. The annihilation channels that are responsible for keeping $Z^\prime$ and $S$ in the thermal bath with the SM are shown in Fig.\,\ref{xy}. We have chosen the model parameters in such a way that the $\lambda_{s\phi}\gtrsim 10^{-4}$ and $g_x$ vary between $(0-1)$. Such a choice of parameters ensures the thermalisation of $Z^\prime$ in the early universe. 
 \noindent DM phenomenology is sensitive to the mass of DM ($M_Z^\prime$), $U(1)_X$ gauge coupling ($g_x$), the mass of the singlet component dominated scalar ($m_S$),  the mixing angles ($\sin\alpha_2, \sin\alpha_3$) and the mass of the triplet scalar ($m_\Delta$).

\begin{figure}
    \centering
    $$
    \includegraphics[scale=0.35]{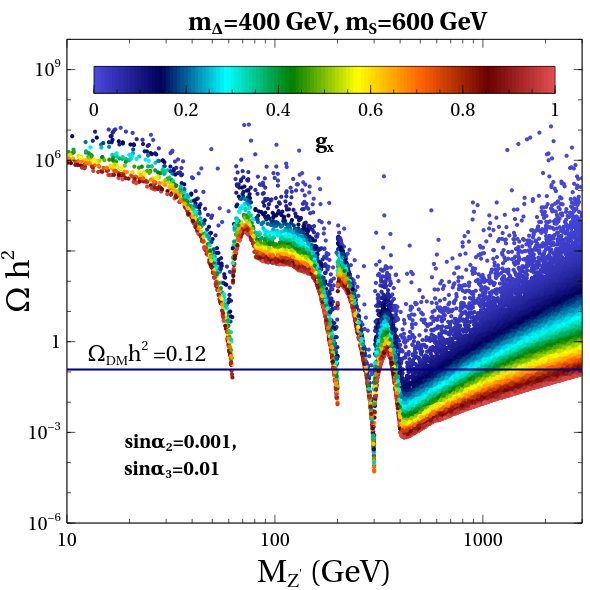}~~~~~~~~~~~
    \includegraphics[scale=0.35]{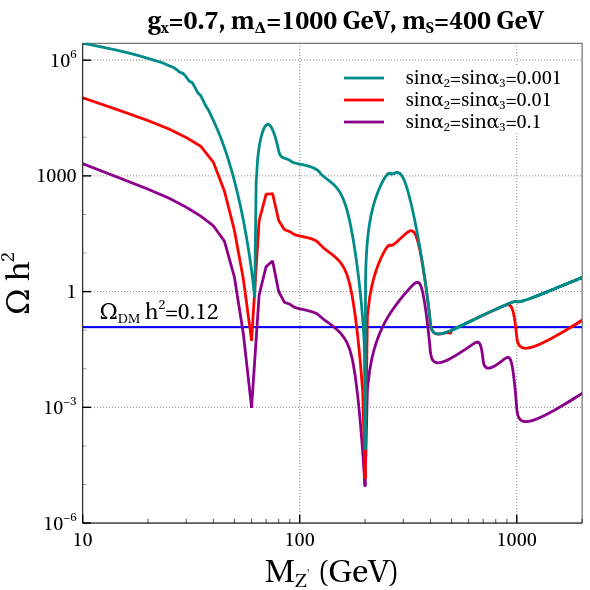}
    $$
    \caption{ Variation of relic abundance with the mass of DM. In the left panel, The value of $m_{\Delta}$ taken is $400$ GeV and $m_S$ is $600$ GeV. The value of mixing angles are $\sin\alpha_{2}=0.001$ and $\sin\alpha_{3}=0.01$. The color bar corresponds to the variation of $g_x$. In the right panel, The value of $m_{\Delta}$ taken is $1000$ GeV and $m_S$ is $400$ GeV. $g_x$ is taken to be $0.7$. The different colored line corresponds to different mixing angles which are mentioned inset.}
    \label{fig:gx_variation}
\end{figure}
The Boltzmann equation governing the co-moving number density of $Z^\prime$ is given by
\begin{equation}
    \frac{dY_{Z^\prime}}{dz}= -\langle \sigma_{Z^\prime Z^\prime \to X X} v_{rel}\rangle \frac{\beta s}{ \mathcal{H} z} (Y^2_{Z^\prime}-{Y^{eq}_{Z^\prime}}^2)
\end{equation}
where $Y_{Z^\prime}=n_{Z^\prime}/s$, the ratio of the number density of $Z^\prime$ and the entropy density of the visible sector and $z=m_S/T$. $\beta$ is defined as $\beta= \frac{g_*^{1/2} (T) \sqrt{g_{\rho} (T)}}{g_{s} (T)}$
where $g_s$ and $g_\rho$ are degrees of freedom associated with entropy and energy densities respectively and $g_{*}^{1/2}= \frac{g_s}{\sqrt{g_\rho}} (1+ \frac{1}{3} \frac{T}{g_s} \frac{dg_s}{dT})$. The Hubble expansion rate $\mathcal{H}$ is given by
\begin{eqnarray}
\mathcal{H}&=& \sqrt{ \frac{ \pi^2 g_{\star\rho}}{90}}\,\frac{T^2}{M_{\rm Pl}}  ~~,\nonumber
\end{eqnarray}
where $g_{\star\rho}$ is the total relativistic degrees of freedom at temperature $T$ contributing to energy density and $M_{\rm Pl}= 2.4 \times 10^{18}$ GeV. $\langle \sigma_{Z^\prime Z^\prime \to X X} v_{rel}\rangle$ corresponds to the thermal average of annihilation cross section of $Z^\prime$ to SM fields and $H_1, A_0, H_2, H^{\pm \pm}, H^{\pm}$.
The relic density $\Omega_{Z^\prime} h^2$ and DD cross sections used in this analysis are calculated with the package {\tt micrOMEGAs} (v5.3.41) \cite{Belanger:2006is} in conjunction with {\tt FeynRules} \cite{Alloul:2013bka}.
In the left panel of Fig.\,\ref{fig:gx_variation}, we show how the relic abundance depends on $M_{Z^\prime}$. The color bar represents the impact of $g_x$. $m_{\Delta}$ and $m_{S}$ have been fixed at  $400~{\rm GeV}$ and $600~{\rm GeV}$ respectively. $\sigma_{Z^\prime Z^\prime \to SM SM}$ is greatly enhanced at
$M_{Z'} = m_{h}/2 , m_{H_1}/2$ and $m_{H_2}/2$, due to s-channel resonances 
(see Fig.\,\ref{xy}). Consequently, relic density falls sharply 
at these values of $M_{Z'}$, as shown in 
Fig.\,\ref{fig:gx_variation}. One can also note that relic density decreases with increasing $g_x$ due to the enhanced annihilation rate of DM which is the usual behaviour of WIMPs.

 In the right panel of Fig.\,\ref{fig:gx_variation}, we show the effect of mixing angles in the variation of relic density as a function of $M_{Z^\prime}$. The masses of $m_S$ and $m_\Delta$ are set at $400$ GeV and $1000$ GeV respectively. The three coloured lines in the figure correspond to three different choices of mixing angles $\sin\alpha_{2}$ and $\sin\alpha_{3}$. With the increment of the mixing angles $\sin\alpha_2$ and $\sin\alpha_3$, the $S$ component increases in the physical scalars $h$ and $H_1$. Consequently, the rate of annihilation of DM via the mediation of these physical scalars increases resulting in lower relic abundance of $Z^\prime$. We see that for every line when $M_Z^\prime$ crosses the mass of $S$, relic density falls due to the opening up of the dominant $Z^\prime Z^\prime \to H_2 H_2 $ annihilation channel. For $\sin\alpha_{2/3}=0.001$, the effect of $Z^\prime Z^\prime \to H^{++} H^{--} /H^{+} H^{-}/H_{1} H_{1}$ channels are not observed, but for increasing mixing angles, the effect of these channels comes into play. The magenta line which corresponds to mixing angles of $0.01$, shows another low relic region just after $M_{Z^\prime}>m_{\Delta}$. But the $\sin\alpha_{2/3}=0.1$ line shows another interesting effect from the annihilation channel $Z^\prime Z^\prime \to H_1~H_{2}$ when $2 M_{Z^\prime}$ becomes larger than  the sum of masses of $H_2$ anf $H_1$ ($m_{\Delta}+m_S$). Additional channels start to contribute to the annihilation of $Z^\prime$ significantly when we increase the mixing angles.

\begin{figure}
    \centering
    $$
    \includegraphics[scale=0.37]{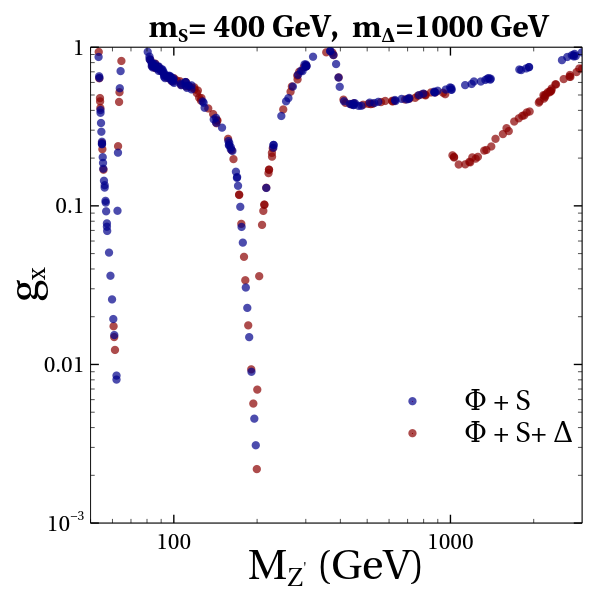}~~~~~~~~~~~~~
    \includegraphics[scale=0.37]{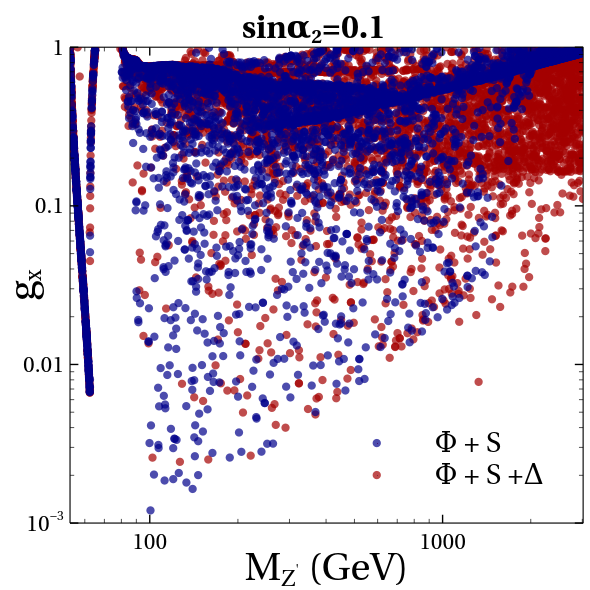}
    $$
    \caption{ Parameter space from relic density constraint in $M_{Z^\prime}$ vs $g_x$ plane. In the left panel, parameter space for correct relic abundance is shown for a fixed mass of $m_S$ and $m_{\Delta}$ (mentioned inset). The dark blue points correspond to the conventional vector dark matter model which has an extra SM singlet in the scalar sector whereas the blue points correspond to our scenario which is extended by another $SU(2)_L$ scalar triplet in comparison to the former one. In the right panel, a similar plot has been shown when $m_S \in (200 - 3000)$ GeV and  $m_\Delta \in (400 - 3000)$ GeV. 
    $\sin\alpha_{2}=0.1$ is considered for both panels. }
    \label{fig:comparison}
\end{figure}

In Fig.\,\ref{fig:comparison}, we show a comparison between the standard $U(1)_X$ extended vector dark matter model and our scenario from a dark matter perspective. The scalar sector of the former model consists of the SM Higgs doublet and a scalar singlet whereas our scenario accommodates another $SU(2)_L$ triplet scalar along with the above-mentioned particles. In Fig.\,\ref{fig:comparison}, in the left panel, allowed parameter space for correct relic abundance is shown for a fixed mass of $m_S$ and $m_{\Delta}$ in $M_{Z^\prime}$ vs $g_x$ plane. The blue points correspond to the standard VDM with only an extra singlet scalar whereas the red points correspond to our scenario (with an additional triplet). The first two resonance funnels in the plot appear due to SM Higgs and singlet scalar resonance which are the same for the two scenarios. However, when $M_{Z^\prime} > 700$ GeV, due to the opening of the extra channel $Z^\prime Z^\prime \to S \Delta$ in our scenario, relic density is satisfied for a lesser value of $g_x$. So the allowed range of $g_x$ for a particular value of $M_{Z^\prime}$ increases. That result is also reflected in the right panel.  The allowed parameter space from relic density constraint is shown by dark blue colored points (dark red points) for the minimal vector dark matter model (our scenario).  We observe that the parameter space is larger for our case allowing lower values of $g_x$ for higher DM mass ranges. This is due to the fact that adding an extra triplet Higgs in the scalar sector increases the parameter space by adding an extra resonance funnel region (see Fig.8) along with providing new annihilation channels. $\sin\alpha_{3}=0.01$ and $\sin\alpha_{2}=0.1$ is considered for both the plots.   

\subsection{Detection Aspects of DM} \label{sec:DD}
\begin{figure}[tbh!]
\centering
\includegraphics[scale=0.4]{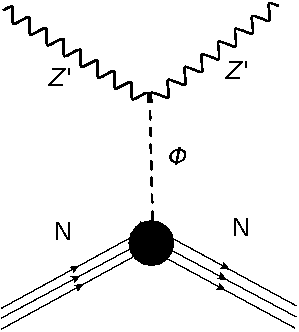}
\caption{ Feynman diagram showing spin-independent elastic scattering of dark matter  $Z^{\prime}$ with nucleons via CP even scalar $\phi$ where $\phi$={$h$,$H_1$,$H_2$}.}
\label{dddd}
\end{figure}

Dark matter detection can be classified broadly into three categories namely direct search, indirect search, and collider search. Direct search experiments measure the cross-section of elastic scattering between DM and nucleons inside the detector material. $Z^\prime$, in the present scenario, scatters off the nucleons via the mixing of $S$ with SM Higgs. The Feynman diagram of the elastic scattering of DM to nucleons is shown in Fig.\,\ref{dddd}. The approximate expression of the cross-section of this t-channel process is given by \cite{Baek:2012se}
\begin{equation}
    \sigma_{Z^\prime N} \simeq \frac{\mu^2_{XN}}{ \pi} (\frac{g_{x} \sin\alpha_{2} \cos\alpha_{2}m_p}{v_d})^2  ( \frac{1}{m^2_h}-\frac{1}{m^2_{S}})^2 f^2_p
    \label{eq:dd}
\end{equation}
where $\mu_{XN}=\frac{M_N M_{Z^\prime}}{M_N+ M_{Z^\prime}}$ is reduced nucleon-DM mass and $f_p = \sum_{q=u,d,s} f^p_q+2/9 (1- \sum_{q=u,d,s} f^p_q) \simeq 0.468$ \cite{Belanger:2008sj}. Here the effect of $H_1$ has been neglected in the formula of the DD cross-section. $H_1$, which is dominantly triplet, does have interaction with the nucleons or the DM  suppressed by small factors involving mixing angles. As a result, the effective DM-nucleon scattering amplitude mediated by $H_1$ is proportional to  $\sin\alpha_1 \sin\alpha_3$. Therefore, in the small $\sin\alpha$ limit, the effect of $H_1$ can be neglected in the DM nucleon scattering cross-section.

\begin{figure}[tbh!]
    \centering
    $$
    \includegraphics[scale=0.35]{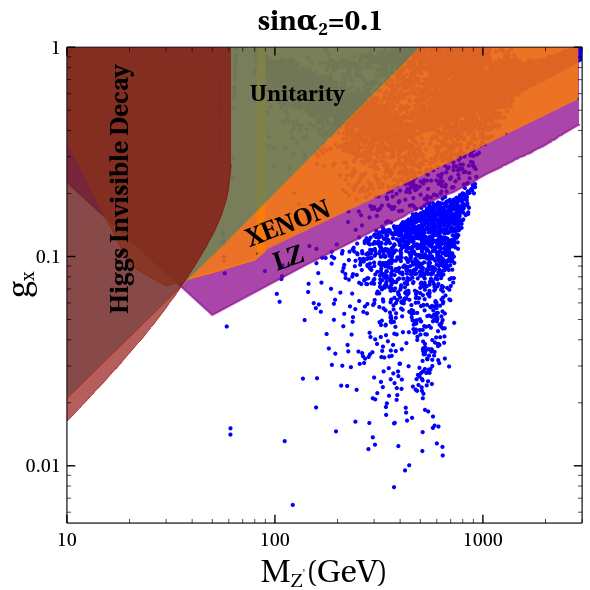}~~~~~~~~~~~~~~~~
    \includegraphics[scale=0.35]{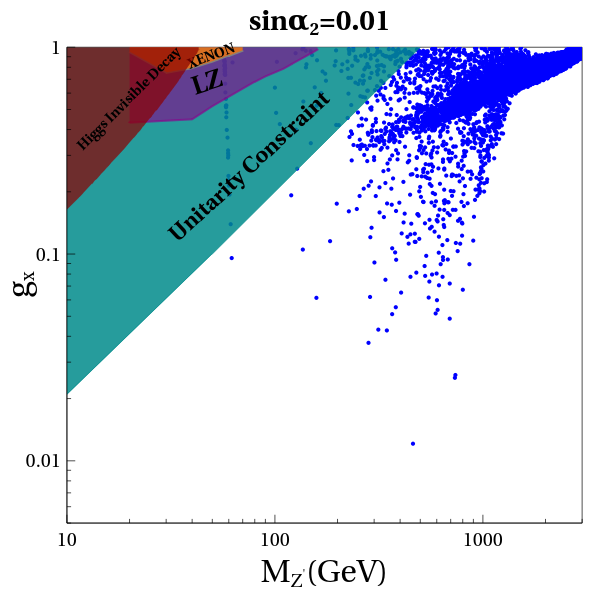}
    $$
    \caption{ Parameter space with all mentioned constraints in  $M_{Z^\prime}$ vs $g_x$ plane for  $\sin\alpha_{2}=0.1$ in the left panel. The coloured region corresponds to exclusion from different constraints. In the right panel, the same plot for $\sin\alpha_{2}=0.01$.}
    \label{fig:dirdet}
\end{figure}

\noindent As given in the Eqn.\,\ref{eq:dd}, the direct search cross-section is sensitive to the parameters $M_{Z^\prime}$, $g_x$, $\sin\alpha_{2}$ and $m_{S}$. The product of $g_x$ and $\sin\alpha_{2}$ act as the effective coupling here. By imposing the collider bound on $\sin\alpha_{2}$, we can have an upper bound on the maximum allowed value of $g_x$ from DD limits. In Fig.\,\ref{fig:dirdet}, we show the bound on $g_x$ for a particular $\sin\alpha_{2}$ over a model parameter space satisfying all other constraints mentioned. In the left panel of Fig.\,\ref{fig:dirdet}, we show the relic density satisfied parameter space in  $M_{Z^\prime}$ vs $g_x$ plane for a range of scalar masses allowed by vacuum stability and unitarity for $\sin\alpha_{2}=0.1$. It is observed that the maximum allowed value of $g_x$ is $\sim 0.3$ in this case from all constraints. The DD cross-section from the LZ experiment gives the most stringent bound and XENON constraints next to it. The other bounds include Higgs invisible decay width and the bound on $v_s$ from unitarity though their effect becomes important for smaller $\sin\alpha_2$. In the right panel of Fig.\,\ref{fig:dirdet}, we show the parameter space for $\sin\alpha_{2}=0.01$ in the same plane. Here one can see that by lowering the value of $\sin\alpha_{2}$ by one order, the constraint on $g_x$ is significantly relaxed from DD. 

Indirect search provides a complementary way to look for DM in the universe. WIMP annihilation in the galactic center, nearby galaxies, or the sun can produce observable fluxes of SM particles (photons, electrons, neutrinos, etc.) which can be detected at various telescopes giving us information about DM distribution. Fermi-LAT, a satellite-based experimental facility, studies cosmic gamma-ray fluxes coming from the galactic centre. It provides the strongest bound on the rate of DM annihilation to photons for DM masses up to few hundreds of GeV \cite{Goodenough:2009gk}. 
 In the present framework, the generation of monoenergetic and diffused photons from DM annihilation is driven by the processes $Z^\prime Z^\prime \to H_2 \to \gamma \gamma $ 
at one loop level and via $ Z^\prime Z^\prime \to H_2 \to W^\pm  W^\pm, H^\pm H^\mp, H^{\pm \pm} H^{\mp \mp}, f {\bar f}$ (with photons emitted from all 
external charged particles) if kinematically possible. However, the process $Z^\prime Z^\prime \to H_2 \to \gamma \gamma $ is 
not only loop-suppressed but also subject to mixing angle suppression. The other channels are solely affected by mixing angle suppression. 
Consequently, due to the combined effects of loop and mixing angle suppression, both the line and diffuse gamma-ray fluxes resulting from dark matter 
annihilation in our model are significantly smaller than the limits indicated by Fermi-LAT data in the GeV mass ranges. Other important constraints as mentioned in \cite{Elor:2015bho} come from different experiments like CMB measurement by Planck, the Fermi detection of photons from the
dwarf galaxies, the positron data from AMS-02, etc. However in our model, as DM annihilates to non-standard Higgses dominantly, so the direct production cross-section of SM particles is well below the observed bounds.

Before we conclude this section, let us very briefly comment on the search prospect of the VDM at the LHC. Unfortunately, the direct production of a $Z^\prime$ pair in pp collision is prohibited by conservation of $U(1)_X$ charge. 
$Z^\prime$ pair can arise from the decay of $H_2$ which can be produced in association with a $Z$ boson at the LHC. In the allowed range of model parameters ($m_{H_2}, g_x, \sin\alpha_{2}$) from the constraints mentioned, the cross-section is significantly small due to $\sin\alpha_{2}$ suppression. Another possible $Z^\prime$ signature can arise from the production of $H^{\pm \pm}$ in association with $H^\mp$ followed by the decays of $H^{\pm \pm} \rightarrow W^\pm W^\pm$ and $H^\mp \rightarrow W^\mp H_2$ and further followed by $H_2 \rightarrow Z^\prime Z^\prime$. This final state would eventually result into {\em two~dilepton} $+$ {\em di-jet}$+${\em missing~energy} signature. This cross-section is also very small due to the tiny $H^\mp \rightarrow W^\mp H_2 $ suppressed branching ratio for allowed values of $m_{H_2}$ at the LHC. It seems that the detection of our VDM at the LHC is very challenging unless a more innovative method is devised.

\section{Possibility of Freeze-in} \label{sec:freezein}
\begin{figure}[tbh]
    \centering
    $$
    \includegraphics[scale=0.4]{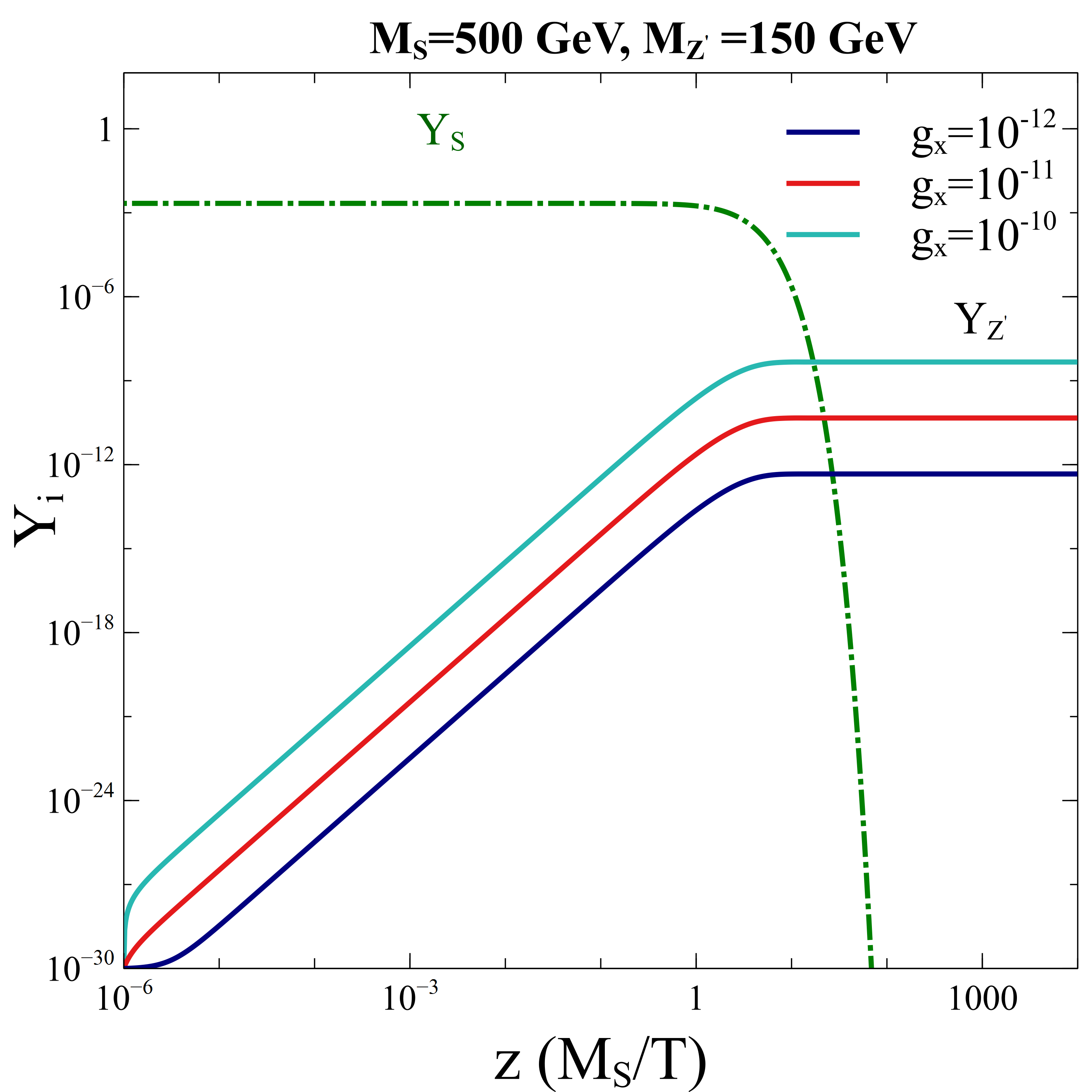}~~~~~~~~~~~
    \includegraphics[scale=0.4]{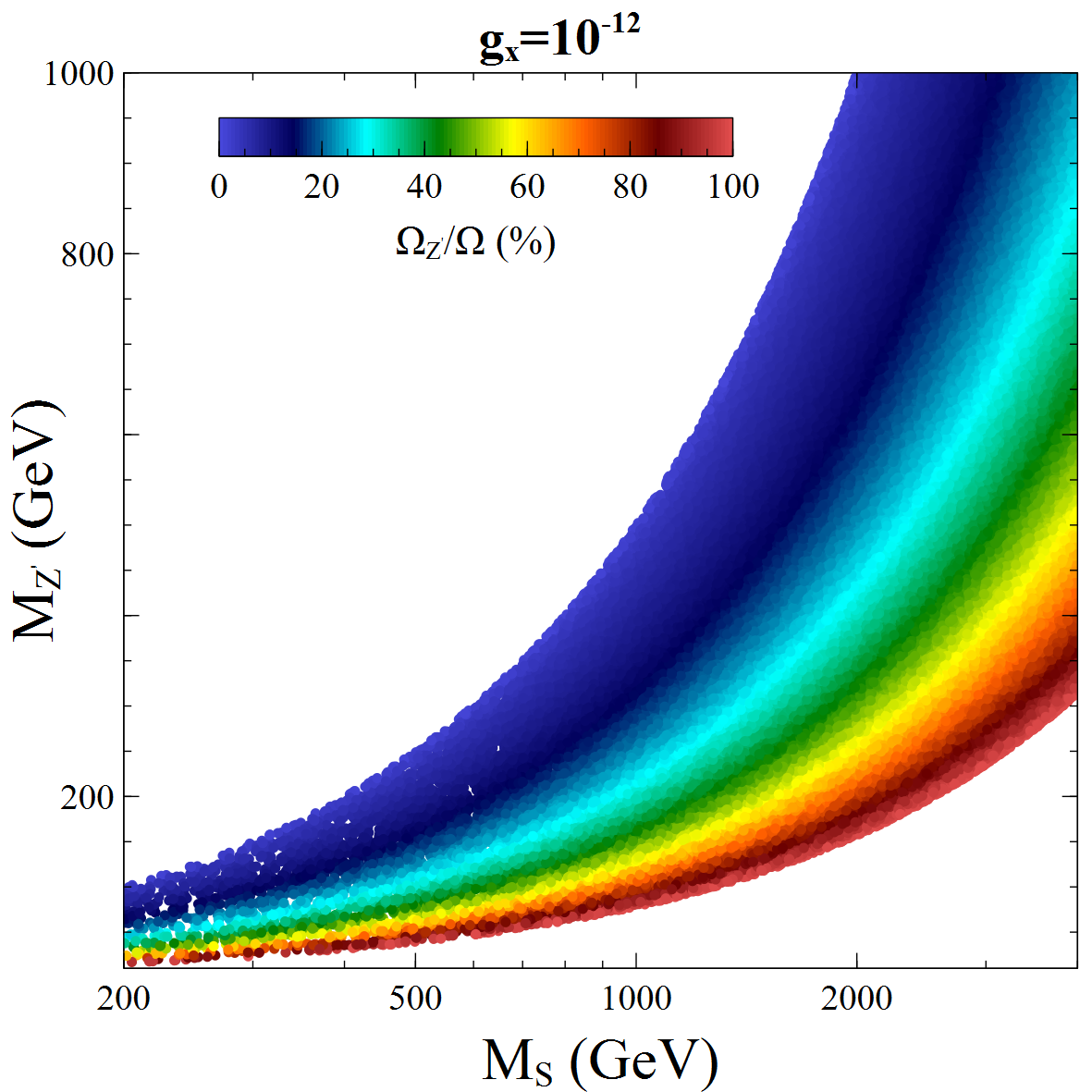}
    $$
    \caption{ Comoving number density evolution of DM and parent particle with z $=(M_S/T)$ in the left panel.  The mass of $ Z^\prime $ and $S$ are mentioned inset. The solid lines correspond to DM number density $Y_{Z^\prime}$ for different values of coupling $g_x$ whereas the dash-dotted line corresponds to the parent particle following the thermal bath. Allowed DM parameter space in $M_S$ vs $M_{Z^\prime}$ plane in the right panel. The color bar corresponds to the contribution of our DM candidate in total relic density in a percentage format.} 
    \label{fig:freezein1}
\end{figure}

In the previous section, we explored a scenario where the DM was initially in thermal equilibrium with the SM particle bath, but at a later time, it froze out as the interaction rate dropped below the expansion rate of the universe ($\mathcal{H}$). Due to such interaction with the SM particles, it can be probed by different DD experiments as we have shown earlier. However, the null results of all the DD experiments motivate us to explore another possibility where the interaction of DM with SM is so feeble that it never reaches the thermal equilibrium in the early universe. Rather, it could be produced from the decay or annihilation of the bath particles. Small value of coupling necessary for such feeble interaction can be naturally generated in a number of ways as shown in Ref.\,\cite{Babu:2023zni} in a flavor model context. Due to such feeble interactions, DM can evade the stringent DD constraints and collider constraints. In this model, $Z^\prime$ can gain its number density in the early universe from the decay of singlet scalar $S$ ($H_2$) or annihilation of $S$ ($H_2$). As $S$ can be thermally produced through its interaction with SM Higgs, if the gauge coupling of $U(1)_X$, $g_x$ is very small, typically in the order of $~O(10^{-12})$, then $Z^\prime$ can never attain thermal equilibrium with the SM bath. With negligible initial abundance, DM gains its number density gradually from the decay of $S$. 
The decay of $S$ is dictated by the vertex factor corresponding to $S {Z^\prime}^\mu Z^\prime_\mu \sim g_x^2 v_S$ whereas the annihilation of $S$ is controlled by the vertex factor of $S^* S {Z^\prime}^\mu Z^\prime_\mu \sim g_x^2$. 
As the annihilation of $S$ is suppressed by a $v_s$ factor with respect to the decay at the vertex factor level, the decay term gives a dominant contribution to DM production.
The mixing of $S$ to other scalars being negligible $S$ can be identified with mass eigenstate $H_2$. To find out the DM number density evolution, we need to solve the Boltzmann equation for the number density of $Z^\prime$. The Boltzmann equation in terms of  DM comoving number density $Y_{Z^\prime}$ is given by
\begin{equation}
    \frac{dY_{Z^\prime}}{dz}= \frac{\langle\Gamma_{S \to Z^\prime Z^\prime} \rangle}{ \mathcal{H} z} Y^{eq}_S (z)
\end{equation}
where $z$ is $\frac{M_S}{T}$, $Y_{Z^\prime}$ is the ratio of the comoving number density of DM to the entropy density of the visible sector and $Y^{eq}_S$ is the equilibrium co-moving number density of $S$ which is given by
\begin{equation}
    Y^{eq}_S= \frac{45}{4 \pi^4} \frac{g_s}{g_{*s}} (M_S/T)^2 K_2 (M_S/T)
\end{equation}
 In the left panel of Fig.\,\ref{fig:freezein1}, we discuss the density evolution of our non-thermal DM candidate $Z^\prime$ by solving the Boltzmann equation. The solid lines correspond to the DM number densities for different values of $U(1)_X$ coupling $g_x$. As the coupling $g_x$ decreases, the decay width of $S$ decreases, therefore the final DM density achieved from the decay of $S$ reduces. The dash-dotted line shows the number density of parent particle $S$ which follows thermal equilibrium. The DM abundance in this current era can be obtained as
 \begin{equation}
     \Omega_{\rm Z^\prime} h^2 = (2.755 \times 10^8) \frac{M_{Z^\prime}}{\rm GeV} Y^{\rm today}_{Z^\prime} 
 \end{equation}
 Now comoving number density of DM $Z^\prime$, $Y_{Z^\prime}$ can be analytically expressed as \cite{Hall:2009bx}
 \begin{equation}
     Y^{today}_{Z^\prime} \approx \frac{135 \, g_S}{8 \pi^3 (1.66) g_{* s} \sqrt{g_{*}} } \frac{M_{PL}}{M^2_S} \, \Gamma_{S \to Z^\prime Z^\prime}
 \end{equation}
 which matches with our solution from the Boltzmann equation.
In the right panel of Fig.\,\ref{fig:freezein1}, we show a region in $M_S$ vs $M_Z^{\prime}$ plane where $Z^\prime$ can contribute to the relic density through the decay of scalar $S$ for $g_x= 10^{-12}$. The color bar corresponds to the percentage of relic density contribution of our DM candidate $Z^\prime$. The region under the allowed parameter space is overabundant whereas the upper region is excluded by kinematics ($2 M_{Z^\prime}\geq m_{S}$).

\section{Conclusion} \label{sec:summ}
To summarise, we have studied the phenomenology of a VDM in a gauged $U(1)_X$ extension of the SM set in a type II seesaw framework. The light neutrino masses are generated via the type-II seesaw mechanism by the introduction of an additional Higgs triplet ($\Delta$). In the context of such a model, we investigate the possibility of a VDM, $Z^\prime$ which is stabilised by an exact dark conjugation symmetry. $Z^\prime$ becomes massive when a complex scalar singlet $S$ acquires VEV breaking the $U(1)_X$ symmetry spontaneously. We studied the possibility of both thermal and non-thermal production of $Z^\prime$ in the early universe. Thermalisation of $Z^\prime$ in the early universe can be achieved for a sizable range of values of $g_x$ ($0-1$) along with a comparatively smaller value of $\lambda_{s\phi}$ ($\mathcal{O}(10^{-4})$), the mixing between SM Higgs and the additional singlet scalar ($S$). This mixing acts as a portal between the dark sector and the visible sector. To begin with, we have estimated the density of $Z^\prime$ in the present epoch by implementing our model in micromega. We have then shown a comparison between the relic density satisfied parameter space of our scenario and that of the $U(1)_X$ extended minimal VDM model. We found out that our scenario allows lower $g_x$ values for larger mass of DM due to presence of newer annihilation channels with respect to the minimal model. The resulting relic density has been compared with available data reported by the PLANCK satellite. We have shown the implications of such a WIMP DM in light of the current limits of DD experiments like XENON and LZ. Due to the presence of the triplet scalar, necessary for neutrino mass generation, the vacuum stability and unitarity constraints on $\lambda$ parameters are restrictive. Our model parameter space satisfies constraints from vacuum stability, perturbative unitarity, $\rho$ parameter, oblique parameters, lepton flavour violation from charged scalars, and Higgs invisible decay. This framework
can also accommodate the existing $2\sigma$ deviation observed in the $h \to Z \gamma$ decay rate over a wide range of parameter space. We have shown that after considering all the aforementioned constraints, the upper limit of $g_x$ is nearly $0.3$ for DM masses in the range $(100 - 1000)$ GeV when $\sin\alpha_2 =0.1$. However, such an upper limit on $g_x$ crucially depends on $\sin\alpha_2$ and becomes weaker for lower values of $\sin\alpha_2$ as evident from Fig.\,\ref{fig:dirdet}. In the end, we briefly discussed the possibility of non-thermal production of $Z^\prime$ from the decay of the complex scalar $S$. For the values of $g_x$ as small as $10^{-12}$, the allowed mass range of DM varies between $(100-300)$ GeV when $Z^\prime$ contributes $100\%$ to the relic density.

\section*{Acknowledgement} \label{sec:ackno}
The authors would like to thank Dilip Kumar Ghosh for various useful discussions during this work. ND would like to thank Pooja Bhattacharjee for useful discussions. ND is funded by CSIR, Government of India, under the NET SRF fellowship scheme with Award file No.09/080(1187)/2021-EMR-I. TJ acknowledges the support from the Science and Engineering Research Board (SERB), Government of India under grant reference no. PDF/2020/001053. The work of DN is supported by the National Research Foundation of Korea (NRF) Research Grant NRF-2019R1A2C3005009 (DN).
\appendix
\section{Appendix} \label{sec:appen}
\subsection{Expression for $\lambda$ parameters }
\label{ss:lambda}
The $\lambda$ parameters in terms of masses of physical scalars and mixing angles can be expressed as
\begin{eqnarray} \label{eq:remaininglam}
\lambda &=& \frac{1}{2 v_{d}^{2}}\Big \{m_{h}^{2} c_{\alpha_{1}}^{2} c_{\alpha_{2}}^{2} + s_{\alpha_{1}}^{2}(m_{H_{1}}^{2} c_{\alpha_{3}}^{2} + m_{H_{2}}^{2} s_{\alpha_{3}}^{2}) + c_{\alpha_{1}}^{2}s_{\alpha_{2}}^{2}(m_{H_{1}}^{2} s_{\alpha_{3}}^{2} + m_{H_{2}}^{2} c_{\alpha_{3}}^{2})\nonumber \\ 
&& - (m_{H_{2}}^{2} - m_{H_{1}}^{2}) c_{\alpha_{1}} s_{\alpha_{1}} s_{\alpha_{2}} s_{2\alpha_{3}}\Big \}, \\
\lambda_{1} &=&  \frac{1}{v_{d} v_{t}} \Big \{m_{h}^{2} c_{\alpha_{1}} s_{\alpha_{1}} c_{\alpha_{2}}^{2} -  m_{H_{1}}^{2}(s_{\alpha_{1}}c_{\alpha_{3}} + c_{\alpha_{1}} s_{\alpha_{2}} s_{\alpha_{3}})(c_{\alpha_{1}}c_{\alpha_{3}} - s_{\alpha_{1}} s_{\alpha_{2}} s_{\alpha_{3}}) \nonumber \\ 
&& + m_{H_{2}}^{2}(s_{\alpha_{1}} s_{\alpha_{2}} c_{\alpha_{3}} + c_{\alpha_{1}}s_{\alpha_{3}})(c_{\alpha_{1}} s_{\alpha_{2}} c_{\alpha_{3}} - s_{\alpha_{1}}s_{\alpha_{3}})\Big \} + 2 \Big \{ \frac{2 m_{H^{\pm}}^{2}}{v_{d}^{2} + 2 v_{t}^{2}} - \frac{m_{A}^{2}}{v_{d}^{2} + 4 v_{t}^{2}}\Big \},  \\
\lambda_{2} &=&  \frac{1}{2 v_{t}^{2}} \Big \{m_{h}^{2} s_{\alpha_{1}}^{2} c_{\alpha_{2}}^{2} +  m_{H_{1}}^{2}(c_{\alpha_{1}}c_{\alpha_{3}} - s_{\alpha_{1}} s_{\alpha_{2}} s_{\alpha_{3}})^{2} + m_{H_{2}}^{2}(c_{\alpha_{1}}s_{\alpha_{3}} + s_{\alpha_{1}} s_{\alpha_{2}} c_{\alpha_{3}})^{2} {\Big \} }\nonumber \\ 
&& + \frac{v_{d}^{2}}{2 v_{t}^{2}} \Big \{ \frac{m_{A}^{2}}{v_{d}^{2} + 4 v_{t}^{2}} - \frac{4 m_{H^{\pm}}^{2}}{v_{d}^{2} + 2 v_{t}^{2}}\Big \} + \frac{m_{H^{\pm\pm}}^{2}}{v_{t}^{2}},  \\
\lambda_3 &=&\frac{1}{v^2_t} \Big\{\frac{2 {v^2_d} m^2_{H^\pm}}{v^2_d+2 v^2_t}-m^2_{H^{\pm \pm}}-\frac{v^2_d m^2_{A}}{v^2_d+4 v^2_t}\Big\}  \\
\lambda_4 &=& \frac{4 m^2_{A}}{v^2_d+ 4 v^2_t}-\frac{4 m^2_{H^{\pm}}}{v^2_d +2 v^2_t}\\
\mu &=& \frac{\sqrt{2} m^2_{A^0} v_t}{v^2_d+ 4 v^2_t}\\
\lambda_{s} &=& \frac{1}{2 v_{s}^{2}}\Big \{m_{h}^{2}s_{\alpha_{2}}^{2} + c_{\alpha_{2}}^{2}(m_{H_{1}}^{2} s_{\alpha_{3}}^{2} + m_{H_{2}}^{2} c_{\alpha_{3}}^{2}) \Big \}, \nonumber \\
\lambda_{s\phi} &=& \frac{c_{\alpha_{2}}}{v_{d} v_{s}}\Big\{(m_{H_{2}}^{2} - m_{H_{1}}^{2}) s_{\alpha_{1}} s_{\alpha_{3}} c_{\alpha_{3}} +  c_{\alpha_{1}} s_{\alpha_{2}}(m_{h}^{2} - m_{H_{1}}^{2} s_{\alpha_{3}}^{2} - m_{H_{2}}^{2} c_{\alpha_{3}}^{2})\Big\},\\
\lambda_{S\Delta} &=& \frac{c_{\alpha_{2}}}{v_{t} v_{s}}\Big\{(m_{H_{1}}^{2} - m_{H_{2}}^{2}) c_{\alpha_{1}} s_{\alpha_{3}} c_{\alpha_{3}} +  s_{\alpha_{1}} s_{\alpha_{2}}(m_{h}^{2} - m_{H_{1}}^{2} s_{\alpha_{3}}^{2} - m_{H_{2}}^{2} c_{\alpha_{3}}^{2})\Big\}.
\end{eqnarray}

\subsection{$h \to \gamma \gamma $ and $h \to Z \gamma $ accessories} 
\label{ss:htozgamma}
The loop functions for the spin-0, spin-$1/2$ and spin-1 particles are given by
\begin{eqnarray}
    A^h_0(\tau) &=& -[\tau-f(\tau)] \tau^{-2}, \\
A^h_{1/2}(\tau) &=& 2[\tau+(\tau-1)f(\tau)] \tau^{-2}, \\
A^h_1 (\tau)&=& -[2\tau^2+3 \tau+3(2\tau-1)f(\tau)] \tau^{-2},
\end{eqnarray}
where 
\begin{align*}
f(\tau) = 
\left\{
\begin{aligned}
    & [\sin^{-1}(\sqrt{\tau})]^2 \quad &  (\tau \leq 1) \\
         &-\frac{1}{4}\Bigg[ \log\Big(\frac{1+\sqrt{1-\tau^{-1}}}{1-\sqrt{1-\tau^{-1}}}\Big)-i \pi \Bigg]^2 \quad & (\tau > 1)  
\end{aligned}
\right.
\end{align*}
 $g_{h f \Bar{f}}$ and $g_{h W^{+} W^{-}}$ corresponds to the relative coupling of mass eigenstate h to the SM Higgs coupling. The relative couplings in terms of our set of free parameters can be written as
 \begin{equation}
 \label{eqn:gff}
     g_{h f \Bar{f}}= \frac{\cos\alpha_1 \cos\alpha_2}{\cos\beta^\prime},~~~~ g_{h W^{+} W^{-}}= \frac{2 v_t \sin\alpha_1 \cos\alpha_2 }{v_d}+ \cos\alpha_1 \cos\alpha_2 
 \end{equation}
Here we can see that when $\sin\alpha_2$ tends to zero, we get back our pure type II seesaw couplings (see \cite{BhupalDev:2013xol}). In the limit, $\sin\alpha_1 \to 0$, $\sin\alpha_2 \to 0$ and $v_t << v_d$, $\cos\beta^\prime \to 0$ and the relative couplings become one hence the SM Higgs coupling are restored. The scalar trilinear couplings $\Tilde{g_{h H^{\pm} H^{\mp}}}$ and $\Tilde{g_{h H^{\pm \pm} H^{\mp \mp}}}$ are given by
\begin{equation}
\Tilde{g}_{h H^{++} H^{--}}= \frac{m_w}{g m^2_{H^{\pm \pm}}} g_{h H^{++} H^{--}}, ~~~\Tilde{g}_{h H^{+} H^{-}}= \frac{m_w}{g m^2_{H^{\pm}}} g_{h H^{+} H^{-}} 
\end{equation}
with $g_{h H^{++} H^{--}}$ and $g_{h H^{+} H^{-}}$ given by
\begin{eqnarray}
 g_{h H^{++} H^{--}}&=& \lambda_{1} v_{d} \cos\alpha_{1} \cos\alpha_{2} + 2 \lambda_2 v_t \sin\alpha_1 \cos\alpha_2 + \lambda_{\Delta S} \, v_s \sin\alpha_2 \nonumber \\
   g_{h H^{+} H^{-}} &=& \frac{1}{2} \Big\{ 
   4 (\lambda_2+\lambda_3) v_t \cos^2\beta^\prime \cos\alpha_2+ 2 \lambda_1 v_t \sin^2\beta^\prime \sin\alpha_1 \cos\alpha_2+ \frac{1}{\sqrt{2}} \lambda_4 v_d \sin2\beta^\prime \sin\alpha_1 \cos\alpha_2 \nonumber \\
   &+& 
   \cos\alpha_1
   \cos\alpha_2 (4 \lambda v_d \sin^2\beta^\prime+
(2 \lambda_1+\lambda_4) v_d \cos^2\beta^\prime+(4 \mu-\sqrt{2} \lambda_4 v_t) \cos\beta^\prime \sin\beta^\prime) \nonumber \\
&+& 2 \sin\alpha_2 (\lambda_{\Delta S} \cos^2\beta^\prime+\lambda_{s\phi} \sin^2\beta^\prime) \Big\}
\end{eqnarray} 
The terms including $\lambda_{\Delta S}$ and $\lambda_{\Delta S}$ come from the extra complex singlet scalar sector and a factor of $\cos\alpha_2$ appears with the rest of the part due to its mixing with the SM doublet. In the limit $\sin\alpha_2 \to 0$, these couplings reduce to Doublet Triplet Higgs model couplings as mentioned in \cite{Arhrib:2011vc}.
The necessary loop functions and couplings for $h \to Z \gamma$ are mentioned below.
The loop factors for particles with different spins are given by
\begin{eqnarray}
A^h_0(\tau_h,\tau_Z)&=& I_1 (\tau_h,\tau_Z) \nonumber\\
A^h_{1/2}(\tau_h,\tau_Z)&=& I_1 (\tau_h,\tau_Z)-I_2 (\tau_h,\tau_Z) \nonumber\\
A^h_1(\tau_h,\tau_Z)&=& 4(3-\tan^2\theta_W) I_2(\tau_h,\tau_Z)+ [(1+2 \tau^{-1}_h) \tan^2\theta_W -(5+2 \tau^{-1}_h)] I_1 (\tau_h,\tau_Z) \nonumber\\   
\end{eqnarray}
where the $I_1$ and $I_2$ are given by
\begin{eqnarray}
    I_1(\tau_h,\tau_Z) &=&\frac{\tau_h \tau_Z}{2 (\tau_h-\tau_Z)} +\frac{\tau^2_h \tau^2_Z}{2 (\tau_h-\tau_Z)^2}[f[\tau^{-1}_h]- f[\tau^{-1}_Z]]+\frac{\tau^2_h \tau_Z}{(\tau_h-\tau_Z)^2}[g(\tau^{-1}_h-g(\tau^{-1}_Z))],\nonumber \\
    I_2(\tau_h,\tau_Z)&=&-\frac{\tau_h \tau_Z}{2 (\tau_h-\tau_Z)} [f[\tau^{-1}_h]- f[\tau^{-1}_Z]]
\end{eqnarray}
Here the function $g(\tau)$ is given by
\begin{align*}
g(\tau) = 
\left\{
    \begin {aligned}
         & \sqrt{\tau^{-1}-1}\sin^{-1}(\sqrt{\tau}) \quad &  (\tau < 1) \\
         &\frac{1}{2} \sqrt{1-\tau^{-1}}\Bigg[ \log\Big(\frac{1+\sqrt{1-\tau^{-1}}}{1-\sqrt{1-\tau^{-1}}}\Big)-i \pi \Bigg]^2 \quad & (\tau \geq 1)                  
    \end{aligned}
\right.
\end{align*}
The function $f(\tau)$, couplings $g_{h f \Bar{f}}$ and $g_{h W^{+} W^{-}}$ are given in Eqn.\,\ref{eqn:gff}. The coupling $g_{Z H^{+} H^{-}}$ and $g_{Z H^{++} H^{--}}$ are given by
\begin{equation}
    g_{Z H^{+} H^{-}}= -\tan\theta_W,~~~
g_{Z H^{++} H^{--}}= 2 \cot2\theta_W    
\end{equation}

\subsection{Unitarty constraints}
\label{appendix:unitarity}
The unitary bounds of the model are extracted from the amplitude matrix ($2 \to 2$ scattering amplitude ) where the basis eigen vector is made of all possible two particle states. Each eigen value of this matrix should be less than $8 \pi$ to satisfy unitarity. The amplitude matrix can be decomposed into
9 charge neutral, 14 singly charged and 9 doubly charged two particle states. We have calculated the conditions following the prescription of Ref.\,\cite{Arhrib:2011uy} which have been mentioned in the main text. However, the amplitude matrix corresponding to the charge-neutral scatterings does not give the closed form of eigenvalues and therefore we solve the matrix numerically. The charge neutral two particle states with identical particles are given by 
\begin{equation*}
    \ket{\Delta^{++}\Delta^{--}},\ket{\Delta^{+} \Delta^{-}},\ket{\phi^{+} \phi^{-}},\ket{g_0 g_0},\ket{\eta_0 \eta_0},\ket{h_0 h_0},\ket{\delta_0 \delta_0},\ket{s_r s_r},\ket{s_i s_i}.
\end{equation*}
The amplitude matrix in the above basis is given by

\begin{equation}
   M= \begin{pmatrix}
 4 (\lambda_2+ \lambda_3) & 2 (\lambda_2+ \lambda_3) & \lambda_1+\lambda_4 & \frac{\lambda
   _1}{\sqrt{2}} & \frac{\lambda1}{\sqrt{2}} & \sqrt{2} \lambda_2 & \sqrt{2} \lambda_2 & \frac{\lambda_{\Delta s}}{\sqrt{2}} &
   \frac{\lambda_{\Delta s}}{\sqrt{2}} \\
 2 (\lambda_2+\lambda_3) & 4 \lambda
   _2+\frac{\lambda_3}{2} & \lambda_1 +\frac{\lambda
   _4}{2} & \frac{2 \lambda_1+\lambda_4}{2 \sqrt{2}} & \frac{2 \lambda_1+\lambda_4}{2 \sqrt{2}} & \sqrt{2} (\lambda
   _2+\lambda_3) & \sqrt{2} (\lambda
   _2+\lambda_3) &
   \frac{\lambda_{\Delta s}}{\sqrt{2}} & \frac{\lambda_{\Delta s}}{\sqrt{2}} \\
 \lambda_1+\lambda_4 & \lambda_1+\frac{\lambda
   _4}{2} & 4 \lambda  & \sqrt{2} \lambda  & \sqrt{2} \lambda  &
   \frac{\lambda_1}{\sqrt{2}} & \frac{\lambda_1}{\sqrt{2}} &
   \frac{\lambda_{\Phi s}}{\sqrt{2}} & \frac{\lambda_{\Phi s}}{\sqrt{2}} \\
 \frac{\lambda_1}{\sqrt{2}} & \frac{2 \lambda_1+\lambda_4}{2 \sqrt{2}} & \sqrt{2} \lambda  & 3 \lambda  & \lambda  &
   \frac{\lambda_1+\lambda_4}{2} & \frac{\lambda_1+\lambda_4}{2} & \frac{\lambda_{\Phi s}}{\sqrt{2}} & \frac{\lambda_{\Phi s}}{\sqrt{2}} \\
\frac{\lambda_1}{\sqrt{2}} & \frac{2 \lambda_1+\lambda_4}{2 \sqrt{2}} & \sqrt{2} \lambda  &  \lambda  &  3 \lambda  &
   \frac{\lambda_1+\lambda_4}{2} & \frac{\lambda_1+\lambda_4}{2} & \frac{\lambda_{\Phi s}}{\sqrt{2}} & \frac{\lambda_{\Phi s}}{\sqrt{2}} \\
 \sqrt{2} \lambda_2 & \sqrt{2} (\lambda_2+\lambda_3) &
   \frac{\lambda_1}{\sqrt{2}} & \frac{\lambda_1+\lambda
   _4}{2} & \frac{\lambda_1+\lambda
   _4}{2} & 3 (\lambda
   _2+\lambda_3) & \lambda_2+\lambda_3 &
   \frac{\lambda_{\Delta s}}{2} & \frac{\lambda_{\Delta s}}{2} \\
 \sqrt{2} \lambda_2 & \sqrt{2} (\lambda_2+\lambda_3) &
   \frac{\lambda_1}{\sqrt{2}} & \frac{\lambda_1+\lambda
   _4}{2} & \frac{\lambda_1+\lambda
   _4}{2} & \lambda
   _2+\lambda_3 & 3 (\lambda_2+\lambda_3) &
   \frac{\lambda_{\Delta s}}{2} & \frac{\lambda_{\Delta s}}{2} \\
 \frac{\lambda_{\Delta s}}{\sqrt{2}} & \frac{\lambda_{\Delta s}}{\sqrt{2}} & \frac{\lambda_{\Phi s}}{\sqrt{2}} & \frac{\lambda
   _{\Phi s}}{2} & \frac{\lambda
   _{\Phi s}}{2} & \frac{\lambda_{\Delta s}}{2} & \frac{\lambda_{\Delta s}}{2} & 3 \lambda_s &
   \lambda_s \\
\frac{\lambda_{\Delta s}}{\sqrt{2}} & \frac{\lambda_{\Delta s}}{\sqrt{2}} & \frac{\lambda_{\Phi s}}{\sqrt{2}} & \frac{\lambda
   _{\Phi s}}{2} & \frac{\lambda
   _{\Phi s}}{2} & \frac{\lambda_{\Delta s}}{2} & \frac{\lambda_{\Delta s}}{2} &  \lambda_s &
  3 \lambda_s\\
\end{pmatrix}
\end{equation}
\subsection{Decay Widths}
The partial decay width of SM Higgs $h$ and singlet scalar $H_2$ to the gauge boson $Z^\prime$ is given by
\begin{eqnarray}
 \Gamma(h \to Z^\prime Z^\prime) &=& \frac{g^2_x \sin^2\alpha_{2}}{8 \pi} \frac{M^2_{Z^\prime}}{m_h} \sqrt{1-\frac{4 M^2_{Z^\prime}}{m^2_h}} (2+\frac{m^4_h}{4 M^4_{Z^\prime}}(1-\frac{2 M^2_{Z^\prime}}{m^2_h})^2) \\
 \Gamma(H_1 \to Z^\prime Z^\prime) &=& \frac{g^2_x \cos^2\alpha_{2}
    \sin^2\alpha_{3}} {8 \pi} \frac{M^2_{Z^\prime}}{m_{H_1}} \sqrt{1-\frac{4 M^2_{Z^\prime}}{m^2_{H_1}}} (2+\frac{m^4_{H_1}}{4 M^4_{Z^\prime}}(1-\frac{2 M^2_{Z^\prime}}{m^2_{H_1}})^2) \\
    \Gamma(H_2 \to Z^\prime Z^\prime) &=& \frac{g^2_x \cos^2\alpha_{2}
    \cos^2\alpha_{3}} {8 \pi} \frac{M^2_{Z^\prime}}{m_{H_2}} \sqrt{1-\frac{4 M^2_{Z^\prime}}{m^2_{H_2}}} (2+\frac{m^4_{H_2}}{4 M^4_{Z^\prime}}(1-\frac{2 M^2_{Z^\prime}}{m^2_{H_2}})^2) 
\end{eqnarray}
\subsection{Interaction Rates}
The thermally averaged decay width of $S$ is given by
\begin{equation}
    \langle\Gamma_{S \to Z^\prime Z^\prime} \rangle = \frac{K_1 (M_S/T)}{K_2 (M_S/T)} \Gamma_{S \to Z^\prime Z^\prime}
\end{equation}
\bibliography{Bibliography.bib}

\begin{thebibliography}{89}
\expandafter\ifx\csname natexlab\endcsname\relax\def\natexlab#1{#1}\fi
\expandafter\ifx\csname bibnamefont\endcsname\relax
  \def\bibnamefont#1{#1}\fi
\expandafter\ifx\csname bibfnamefont\endcsname\relax
  \def\bibfnamefont#1{#1}\fi
\expandafter\ifx\csname citenamefont\endcsname\relax
  \def\citenamefont#1{#1}\fi
\expandafter\ifx\csname url\endcsname\relax
  \def\url#1{\texttt{#1}}\fi
\expandafter\ifx\csname urlprefix\endcsname\relax\def\urlprefix{URL }\fi
\providecommand{\bibinfo}[2]{#2}
\providecommand{\eprint}[2][]{\url{#2}}

\bibitem[{\citenamefont{Zwicky}(1933)}]{Zwicky:1933gu}
\bibinfo{author}{\bibfnamefont{F.}~\bibnamefont{Zwicky}},
  \bibinfo{journal}{Helv. Phys. Acta} \textbf{\bibinfo{volume}{6}},
  \bibinfo{pages}{110} (\bibinfo{year}{1933}).

\bibitem[{\citenamefont{Rubin and Ford}(1970)}]{Rubin:1970zza}
\bibinfo{author}{\bibfnamefont{V.~C.} \bibnamefont{Rubin}} \bibnamefont{and}
  \bibinfo{author}{\bibfnamefont{W.~K.} \bibnamefont{Ford},
  \bibfnamefont{Jr.}}, \bibinfo{journal}{Astrophys. J.}
  \textbf{\bibinfo{volume}{159}}, \bibinfo{pages}{379} (\bibinfo{year}{1970}).

\bibitem[{\citenamefont{Clowe et~al.}(2006)\citenamefont{Clowe, Bradac,
  Gonzalez, Markevitch, Randall, Jones, and Zaritsky}}]{Clowe:2006eq}
\bibinfo{author}{\bibfnamefont{D.}~\bibnamefont{Clowe}},
  \bibinfo{author}{\bibfnamefont{M.}~\bibnamefont{Bradac}},
  \bibinfo{author}{\bibfnamefont{A.~H.} \bibnamefont{Gonzalez}},
  \bibinfo{author}{\bibfnamefont{M.}~\bibnamefont{Markevitch}},
  \bibinfo{author}{\bibfnamefont{S.~W.} \bibnamefont{Randall}},
  \bibinfo{author}{\bibfnamefont{C.}~\bibnamefont{Jones}}, \bibnamefont{and}
  \bibinfo{author}{\bibfnamefont{D.}~\bibnamefont{Zaritsky}},
  \bibinfo{journal}{Astrophys. J. Lett.} \textbf{\bibinfo{volume}{648}},
  \bibinfo{pages}{L109} (\bibinfo{year}{2006}), \eprint{astro-ph/0608407}.

\bibitem[{\citenamefont{Spergel et~al.}(2007)}]{WMAP:2006bqn}
\bibinfo{author}{\bibfnamefont{D.~N.} \bibnamefont{Spergel}}
  \bibnamefont{et~al.} (\bibinfo{collaboration}{WMAP}),
  \bibinfo{journal}{Astrophys. J. Suppl.} \textbf{\bibinfo{volume}{170}},
  \bibinfo{pages}{377} (\bibinfo{year}{2007}), \eprint{astro-ph/0603449}.

\bibitem[{\citenamefont{Roszkowski et~al.}(2018)\citenamefont{Roszkowski,
  Sessolo, and Trojanowski}}]{Roszkowski:2017nbc}
\bibinfo{author}{\bibfnamefont{L.}~\bibnamefont{Roszkowski}},
  \bibinfo{author}{\bibfnamefont{E.~M.} \bibnamefont{Sessolo}},
  \bibnamefont{and}
  \bibinfo{author}{\bibfnamefont{S.}~\bibnamefont{Trojanowski}},
  \bibinfo{journal}{Rept. Prog. Phys.} \textbf{\bibinfo{volume}{81}},
  \bibinfo{pages}{066201} (\bibinfo{year}{2018}), \eprint{1707.06277}.

\bibitem[{\citenamefont{Aghanim et~al.}(2020)}]{Planck:2018vyg}
\bibinfo{author}{\bibfnamefont{N.}~\bibnamefont{Aghanim}} \bibnamefont{et~al.}
  (\bibinfo{collaboration}{Planck}), \bibinfo{journal}{Astron. Astrophys.}
  \textbf{\bibinfo{volume}{641}}, \bibinfo{pages}{A6} (\bibinfo{year}{2020}),
  \bibinfo{note}{[Erratum: Astron.Astrophys. 652, C4 (2021)]},
  \eprint{1807.06209}.

\bibitem[{\citenamefont{Bertone et~al.}(2005)\citenamefont{Bertone, Hooper, and
  Silk}}]{Bertone:2004pz}
\bibinfo{author}{\bibfnamefont{G.}~\bibnamefont{Bertone}},
  \bibinfo{author}{\bibfnamefont{D.}~\bibnamefont{Hooper}}, \bibnamefont{and}
  \bibinfo{author}{\bibfnamefont{J.}~\bibnamefont{Silk}},
  \bibinfo{journal}{Phys. Rept.} \textbf{\bibinfo{volume}{405}},
  \bibinfo{pages}{279} (\bibinfo{year}{2005}), \eprint{hep-ph/0404175}.

\bibitem[{\citenamefont{Hu and Dodelson}(2002)}]{Hu:2001bc}
\bibinfo{author}{\bibfnamefont{W.}~\bibnamefont{Hu}} \bibnamefont{and}
  \bibinfo{author}{\bibfnamefont{S.}~\bibnamefont{Dodelson}},
  \bibinfo{journal}{Ann. Rev. Astron. Astrophys.}
  \textbf{\bibinfo{volume}{40}}, \bibinfo{pages}{171} (\bibinfo{year}{2002}),
  \eprint{astro-ph/0110414}.

\bibitem[{\citenamefont{Kolb and Turner}(1990)}]{Kolb:1990vq}
\bibinfo{author}{\bibfnamefont{E.~W.} \bibnamefont{Kolb}} \bibnamefont{and}
  \bibinfo{author}{\bibfnamefont{M.~S.} \bibnamefont{Turner}},
  \emph{\bibinfo{title}{{The Early Universe}}}, vol.~\bibinfo{volume}{69}
  (\bibinfo{year}{1990}), ISBN \bibinfo{isbn}{978-0-201-62674-2}.

\bibitem[{\citenamefont{Hall et~al.}(2010)\citenamefont{Hall, Jedamzik,
  March-Russell, and West}}]{Hall:2009bx}
\bibinfo{author}{\bibfnamefont{L.~J.} \bibnamefont{Hall}},
  \bibinfo{author}{\bibfnamefont{K.}~\bibnamefont{Jedamzik}},
  \bibinfo{author}{\bibfnamefont{J.}~\bibnamefont{March-Russell}},
  \bibnamefont{and} \bibinfo{author}{\bibfnamefont{S.~M.} \bibnamefont{West}},
  \bibinfo{journal}{JHEP} \textbf{\bibinfo{volume}{03}}, \bibinfo{pages}{080}
  (\bibinfo{year}{2010}), \eprint{0911.1120}.

\bibitem[{\citenamefont{Hochberg et~al.}(2014)\citenamefont{Hochberg, Kuflik,
  Volansky, and Wacker}}]{PhysRevLett.113.171301}
\bibinfo{author}{\bibfnamefont{Y.}~\bibnamefont{Hochberg}},
  \bibinfo{author}{\bibfnamefont{E.}~\bibnamefont{Kuflik}},
  \bibinfo{author}{\bibfnamefont{T.}~\bibnamefont{Volansky}}, \bibnamefont{and}
  \bibinfo{author}{\bibfnamefont{J.~G.} \bibnamefont{Wacker}},
  \bibinfo{journal}{Phys. Rev. Lett.} \textbf{\bibinfo{volume}{113}},
  \bibinfo{pages}{171301} (\bibinfo{year}{2014}),
  \urlprefix\url{https://link.aps.org/doi/10.1103/PhysRevLett.113.171301}.

\bibitem[{\citenamefont{Dimopoulos et~al.}(1990)\citenamefont{Dimopoulos,
  Eichler, Esmailzadeh, and Starkman}}]{PhysRevD.41.2388}
\bibinfo{author}{\bibfnamefont{S.}~\bibnamefont{Dimopoulos}},
  \bibinfo{author}{\bibfnamefont{D.}~\bibnamefont{Eichler}},
  \bibinfo{author}{\bibfnamefont{R.}~\bibnamefont{Esmailzadeh}},
  \bibnamefont{and} \bibinfo{author}{\bibfnamefont{G.~D.}
  \bibnamefont{Starkman}}, \bibinfo{journal}{Phys. Rev. D}
  \textbf{\bibinfo{volume}{41}}, \bibinfo{pages}{2388} (\bibinfo{year}{1990}),
  \urlprefix\url{https://link.aps.org/doi/10.1103/PhysRevD.41.2388}.

\bibitem[{\citenamefont{Feng et~al.}(2003)\citenamefont{Feng, Rajaraman, and
  Takayama}}]{Feng:2003xh}
\bibinfo{author}{\bibfnamefont{J.~L.} \bibnamefont{Feng}},
  \bibinfo{author}{\bibfnamefont{A.}~\bibnamefont{Rajaraman}},
  \bibnamefont{and} \bibinfo{author}{\bibfnamefont{F.}~\bibnamefont{Takayama}},
  \bibinfo{journal}{Phys. Rev. Lett.} \textbf{\bibinfo{volume}{91}},
  \bibinfo{pages}{011302} (\bibinfo{year}{2003}), \eprint{hep-ph/0302215}.

\bibitem[{\citenamefont{Pospelov
  et~al.}(2008{\natexlab{a}})\citenamefont{Pospelov, Ritz, and
  Voloshin}}]{Pospelov:2008jk}
\bibinfo{author}{\bibfnamefont{M.}~\bibnamefont{Pospelov}},
  \bibinfo{author}{\bibfnamefont{A.}~\bibnamefont{Ritz}}, \bibnamefont{and}
  \bibinfo{author}{\bibfnamefont{M.~B.} \bibnamefont{Voloshin}},
  \bibinfo{journal}{Phys. Rev. D} \textbf{\bibinfo{volume}{78}},
  \bibinfo{pages}{115012} (\bibinfo{year}{2008}{\natexlab{a}}),
  \eprint{0807.3279}.

\bibitem[{\citenamefont{Pospelov
  et~al.}(2008{\natexlab{b}})\citenamefont{Pospelov, Ritz, and
  Voloshin}}]{Pospelov:2007mp}
\bibinfo{author}{\bibfnamefont{M.}~\bibnamefont{Pospelov}},
  \bibinfo{author}{\bibfnamefont{A.}~\bibnamefont{Ritz}}, \bibnamefont{and}
  \bibinfo{author}{\bibfnamefont{M.~B.} \bibnamefont{Voloshin}},
  \bibinfo{journal}{Phys. Lett. B} \textbf{\bibinfo{volume}{662}},
  \bibinfo{pages}{53} (\bibinfo{year}{2008}{\natexlab{b}}), \eprint{0711.4866}.

\bibitem[{\citenamefont{D'Agnolo et~al.}(2021)\citenamefont{D'Agnolo, Liu,
  Ruderman, and Wang}}]{DAgnolo:2020mpt}
\bibinfo{author}{\bibfnamefont{R.~T.} \bibnamefont{D'Agnolo}},
  \bibinfo{author}{\bibfnamefont{D.}~\bibnamefont{Liu}},
  \bibinfo{author}{\bibfnamefont{J.~T.} \bibnamefont{Ruderman}},
  \bibnamefont{and} \bibinfo{author}{\bibfnamefont{P.-J.} \bibnamefont{Wang}},
  \bibinfo{journal}{JHEP} \textbf{\bibinfo{volume}{06}}, \bibinfo{pages}{103}
  (\bibinfo{year}{2021}), \eprint{2012.11766}.

\bibitem[{\citenamefont{Akerib et~al.}(2017)}]{LUX:2016ggv}
\bibinfo{author}{\bibfnamefont{D.~S.} \bibnamefont{Akerib}}
  \bibnamefont{et~al.} (\bibinfo{collaboration}{LUX}), \bibinfo{journal}{Phys.
  Rev. Lett.} \textbf{\bibinfo{volume}{118}}, \bibinfo{pages}{021303}
  (\bibinfo{year}{2017}), \eprint{1608.07648}.

\bibitem[{\citenamefont{Aprile et~al.}(2018)}]{XENON:2018voc}
\bibinfo{author}{\bibfnamefont{E.}~\bibnamefont{Aprile}} \bibnamefont{et~al.}
  (\bibinfo{collaboration}{XENON}), \bibinfo{journal}{Phys. Rev. Lett.}
  \textbf{\bibinfo{volume}{121}}, \bibinfo{pages}{111302}
  (\bibinfo{year}{2018}), \eprint{1805.12562}.

\bibitem[{\citenamefont{Aprile et~al.}(2023)}]{XENON:2023cxc}
\bibinfo{author}{\bibfnamefont{E.}~\bibnamefont{Aprile}} \bibnamefont{et~al.}
  (\bibinfo{collaboration}{XENON}), \bibinfo{journal}{Phys. Rev. Lett.}
  \textbf{\bibinfo{volume}{131}}, \bibinfo{pages}{041003}
  (\bibinfo{year}{2023}), \eprint{2303.14729}.

\bibitem[{\citenamefont{Tan et~al.}(2016)}]{PandaX-II:2016vec}
\bibinfo{author}{\bibfnamefont{A.}~\bibnamefont{Tan}} \bibnamefont{et~al.}
  (\bibinfo{collaboration}{PandaX-II}), \bibinfo{journal}{Phys. Rev. Lett.}
  \textbf{\bibinfo{volume}{117}}, \bibinfo{pages}{121303}
  (\bibinfo{year}{2016}), \eprint{1607.07400}.

\bibitem[{\citenamefont{Cui et~al.}(2017)}]{PandaX-II:2017hlx}
\bibinfo{author}{\bibfnamefont{X.}~\bibnamefont{Cui}} \bibnamefont{et~al.}
  (\bibinfo{collaboration}{PandaX-II}), \bibinfo{journal}{Phys. Rev. Lett.}
  \textbf{\bibinfo{volume}{119}}, \bibinfo{pages}{181302}
  (\bibinfo{year}{2017}), \eprint{1708.06917}.

\bibitem[{\citenamefont{Aalbers et~al.}(2023)}]{LZ:2022lsv}
\bibinfo{author}{\bibfnamefont{J.}~\bibnamefont{Aalbers}} \bibnamefont{et~al.}
  (\bibinfo{collaboration}{LZ}), \bibinfo{journal}{Phys. Rev. Lett.}
  \textbf{\bibinfo{volume}{131}}, \bibinfo{pages}{041002}
  (\bibinfo{year}{2023}), \eprint{2207.03764}.

\bibitem[{\citenamefont{Baek et~al.}(2013)\citenamefont{Baek, Ko, Park, and
  Senaha}}]{Baek:2012se}
\bibinfo{author}{\bibfnamefont{S.}~\bibnamefont{Baek}},
  \bibinfo{author}{\bibfnamefont{P.}~\bibnamefont{Ko}},
  \bibinfo{author}{\bibfnamefont{W.-I.} \bibnamefont{Park}}, \bibnamefont{and}
  \bibinfo{author}{\bibfnamefont{E.}~\bibnamefont{Senaha}},
  \bibinfo{journal}{JHEP} \textbf{\bibinfo{volume}{05}}, \bibinfo{pages}{036}
  (\bibinfo{year}{2013}), \eprint{1212.2131}.

\bibitem[{\citenamefont{Farzan and Akbarieh}(2012)}]{Farzan:2012hh}
\bibinfo{author}{\bibfnamefont{Y.}~\bibnamefont{Farzan}} \bibnamefont{and}
  \bibinfo{author}{\bibfnamefont{A.~R.} \bibnamefont{Akbarieh}},
  \bibinfo{journal}{JCAP} \textbf{\bibinfo{volume}{10}}, \bibinfo{pages}{026}
  (\bibinfo{year}{2012}), \eprint{1207.4272}.

\bibitem[{\citenamefont{Baouche et~al.}(2021)\citenamefont{Baouche, Ahriche,
  Faisel, and Nasri}}]{Baouche:2021wwa}
\bibinfo{author}{\bibfnamefont{N.}~\bibnamefont{Baouche}},
  \bibinfo{author}{\bibfnamefont{A.}~\bibnamefont{Ahriche}},
  \bibinfo{author}{\bibfnamefont{G.}~\bibnamefont{Faisel}}, \bibnamefont{and}
  \bibinfo{author}{\bibfnamefont{S.}~\bibnamefont{Nasri}},
  \bibinfo{journal}{Phys. Rev. D} \textbf{\bibinfo{volume}{104}},
  \bibinfo{pages}{075022} (\bibinfo{year}{2021}), \eprint{2105.14387}.

\bibitem[{\citenamefont{Nomura et~al.}(2021)\citenamefont{Nomura, Okada, and
  Yun}}]{Nomura:2020zlm}
\bibinfo{author}{\bibfnamefont{T.}~\bibnamefont{Nomura}},
  \bibinfo{author}{\bibfnamefont{H.}~\bibnamefont{Okada}}, \bibnamefont{and}
  \bibinfo{author}{\bibfnamefont{S.}~\bibnamefont{Yun}},
  \bibinfo{journal}{JHEP} \textbf{\bibinfo{volume}{06}}, \bibinfo{pages}{122}
  (\bibinfo{year}{2021}), \eprint{2012.11377}.

\bibitem[{\citenamefont{Babu et~al.}(2022)\citenamefont{Babu, Jana, and
  Thapa}}]{Babu:2021hef}
\bibinfo{author}{\bibfnamefont{K.~S.} \bibnamefont{Babu}},
  \bibinfo{author}{\bibfnamefont{S.}~\bibnamefont{Jana}}, \bibnamefont{and}
  \bibinfo{author}{\bibfnamefont{A.}~\bibnamefont{Thapa}},
  \bibinfo{journal}{JHEP} \textbf{\bibinfo{volume}{02}}, \bibinfo{pages}{051}
  (\bibinfo{year}{2022}), \eprint{2112.12771}.

\bibitem[{\citenamefont{Barman et~al.}(2017)\citenamefont{Barman, Bhattacharya,
  Patra, and Chakrabortty}}]{Barman:2017yzr}
\bibinfo{author}{\bibfnamefont{B.}~\bibnamefont{Barman}},
  \bibinfo{author}{\bibfnamefont{S.}~\bibnamefont{Bhattacharya}},
  \bibinfo{author}{\bibfnamefont{S.~K.} \bibnamefont{Patra}}, \bibnamefont{and}
  \bibinfo{author}{\bibfnamefont{J.}~\bibnamefont{Chakrabortty}},
  \bibinfo{journal}{JCAP} \textbf{\bibinfo{volume}{12}}, \bibinfo{pages}{021}
  (\bibinfo{year}{2017}), \eprint{1704.04945}.

\bibitem[{\citenamefont{Duch et~al.}(2015)\citenamefont{Duch, Grzadkowski, and
  McGarrie}}]{Duch:2015jta}
\bibinfo{author}{\bibfnamefont{M.}~\bibnamefont{Duch}},
  \bibinfo{author}{\bibfnamefont{B.}~\bibnamefont{Grzadkowski}},
  \bibnamefont{and} \bibinfo{author}{\bibfnamefont{M.}~\bibnamefont{McGarrie}},
  \bibinfo{journal}{JHEP} \textbf{\bibinfo{volume}{09}}, \bibinfo{pages}{162}
  (\bibinfo{year}{2015}), \eprint{1506.08805}.

\bibitem[{\citenamefont{Amiri et~al.}(2022)\citenamefont{Amiri,
  D\'\i{}az~S\'aez, and Ghorbani}}]{Amiri:2022cbv}
\bibinfo{author}{\bibfnamefont{A.}~\bibnamefont{Amiri}},
  \bibinfo{author}{\bibfnamefont{B.}~\bibnamefont{D\'\i{}az~S\'aez}},
  \bibnamefont{and} \bibinfo{author}{\bibfnamefont{K.}~\bibnamefont{Ghorbani}}
  (\bibinfo{year}{2022}), \eprint{2209.11723}.

\bibitem[{\citenamefont{Group et~al.}(2020)\citenamefont{Group, Zyla, Barnett,
  Beringer, Dahl, Dwyer, Groom, Lin, Lugovsky, Pianori
  et~al.}}]{10.1093/ptep/ptaa104}
\bibinfo{author}{\bibfnamefont{P.~D.} \bibnamefont{Group}},
  \bibinfo{author}{\bibfnamefont{P.~A.} \bibnamefont{Zyla}},
  \bibinfo{author}{\bibfnamefont{R.~M.} \bibnamefont{Barnett}},
  \bibinfo{author}{\bibfnamefont{J.}~\bibnamefont{Beringer}},
  \bibinfo{author}{\bibfnamefont{O.}~\bibnamefont{Dahl}},
  \bibinfo{author}{\bibfnamefont{D.~A.} \bibnamefont{Dwyer}},
  \bibinfo{author}{\bibfnamefont{D.~E.} \bibnamefont{Groom}},
  \bibinfo{author}{\bibfnamefont{C.~J.} \bibnamefont{Lin}},
  \bibinfo{author}{\bibfnamefont{K.~S.} \bibnamefont{Lugovsky}},
  \bibinfo{author}{\bibfnamefont{E.}~\bibnamefont{Pianori}},
  \bibnamefont{et~al.}, \bibinfo{journal}{Progress of Theoretical and
  Experimental Physics} \textbf{\bibinfo{volume}{2020}},
  \bibinfo{pages}{083C01} (\bibinfo{year}{2020}), ISSN
  \bibinfo{issn}{2050-3911},
  \eprint{https://academic.oup.com/ptep/article-pdf/2020/8/083C01/34673722/ptaa104.pdf},
  \urlprefix\url{https://doi.org/10.1093/ptep/ptaa104}.

\bibitem[{\citenamefont{Mohapatra et~al.}(2007)}]{Mohapatra:2005wg}
\bibinfo{author}{\bibfnamefont{R.~N.} \bibnamefont{Mohapatra}}
  \bibnamefont{et~al.}, \bibinfo{journal}{Rept. Prog. Phys.}
  \textbf{\bibinfo{volume}{70}}, \bibinfo{pages}{1757} (\bibinfo{year}{2007}),
  \eprint{hep-ph/0510213}.

\bibitem[{\citenamefont{Minkowski}(1977)}]{Minkowski:1977sc}
\bibinfo{author}{\bibfnamefont{P.}~\bibnamefont{Minkowski}},
  \bibinfo{journal}{Phys. Lett. B} \textbf{\bibinfo{volume}{67}},
  \bibinfo{pages}{421} (\bibinfo{year}{1977}).

\bibitem[{\citenamefont{Yanagida}(1979)}]{Yanagida:1979as}
\bibinfo{author}{\bibfnamefont{T.}~\bibnamefont{Yanagida}},
  \bibinfo{journal}{Conf. Proc. C} \textbf{\bibinfo{volume}{7902131}},
  \bibinfo{pages}{95} (\bibinfo{year}{1979}).

\bibitem[{\citenamefont{Gell-Mann et~al.}(1979)\citenamefont{Gell-Mann, Ramond,
  and Slansky}}]{Gell-Mann:1979vob}
\bibinfo{author}{\bibfnamefont{M.}~\bibnamefont{Gell-Mann}},
  \bibinfo{author}{\bibfnamefont{P.}~\bibnamefont{Ramond}}, \bibnamefont{and}
  \bibinfo{author}{\bibfnamefont{R.}~\bibnamefont{Slansky}},
  \bibinfo{journal}{Conf. Proc. C} \textbf{\bibinfo{volume}{790927}},
  \bibinfo{pages}{315} (\bibinfo{year}{1979}), \eprint{1306.4669}.

\bibitem[{\citenamefont{Mohapatra and Senjanovic}(1980)}]{Mohapatra:1979ia}
\bibinfo{author}{\bibfnamefont{R.~N.} \bibnamefont{Mohapatra}}
  \bibnamefont{and}
  \bibinfo{author}{\bibfnamefont{G.}~\bibnamefont{Senjanovic}},
  \bibinfo{journal}{Phys. Rev. Lett.} \textbf{\bibinfo{volume}{44}},
  \bibinfo{pages}{912} (\bibinfo{year}{1980}).

\bibitem[{\citenamefont{Magg and Wetterich}(1980)}]{MAGG198061}
\bibinfo{author}{\bibfnamefont{M.}~\bibnamefont{Magg}} \bibnamefont{and}
  \bibinfo{author}{\bibfnamefont{C.}~\bibnamefont{Wetterich}},
  \bibinfo{journal}{Physics Letters B} \textbf{\bibinfo{volume}{94}},
  \bibinfo{pages}{61} (\bibinfo{year}{1980}), ISSN \bibinfo{issn}{0370-2693},
  \urlprefix\url{https://www.sciencedirect.com/science/article/pii/0370269380908254}.

\bibitem[{\citenamefont{Lazarides et~al.}(1981)\citenamefont{Lazarides, Shafi,
  and Wetterich}}]{LAZARIDES1981287}
\bibinfo{author}{\bibfnamefont{G.}~\bibnamefont{Lazarides}},
  \bibinfo{author}{\bibfnamefont{Q.}~\bibnamefont{Shafi}}, \bibnamefont{and}
  \bibinfo{author}{\bibfnamefont{C.}~\bibnamefont{Wetterich}},
  \bibinfo{journal}{Nuclear Physics B} \textbf{\bibinfo{volume}{181}},
  \bibinfo{pages}{287} (\bibinfo{year}{1981}), ISSN \bibinfo{issn}{0550-3213},
  \urlprefix\url{https://www.sciencedirect.com/science/article/pii/0550321381903540}.

\bibitem[{\citenamefont{Mohapatra and Senjanovi\ifmmode~\acute{c}\else
  \'{c}\fi{}}(1981)}]{PhysRevD.23.165}
\bibinfo{author}{\bibfnamefont{R.~N.} \bibnamefont{Mohapatra}}
  \bibnamefont{and}
  \bibinfo{author}{\bibfnamefont{G.}~\bibnamefont{Senjanovi\ifmmode~\acute{c}\else
  \'{c}\fi{}}}, \bibinfo{journal}{Phys. Rev. D} \textbf{\bibinfo{volume}{23}},
  \bibinfo{pages}{165} (\bibinfo{year}{1981}),
  \urlprefix\url{https://link.aps.org/doi/10.1103/PhysRevD.23.165}.

\bibitem[{\citenamefont{Schechter and Valle}(1980)}]{PhysRevD.22.2227}
\bibinfo{author}{\bibfnamefont{J.}~\bibnamefont{Schechter}} \bibnamefont{and}
  \bibinfo{author}{\bibfnamefont{J.~W.~F.} \bibnamefont{Valle}},
  \bibinfo{journal}{Phys. Rev. D} \textbf{\bibinfo{volume}{22}},
  \bibinfo{pages}{2227} (\bibinfo{year}{1980}),
  \urlprefix\url{https://link.aps.org/doi/10.1103/PhysRevD.22.2227}.

\bibitem[{\citenamefont{Cheng and Li}(1980)}]{PhysRevD.22.2860}
\bibinfo{author}{\bibfnamefont{T.~P.} \bibnamefont{Cheng}} \bibnamefont{and}
  \bibinfo{author}{\bibfnamefont{L.-F.} \bibnamefont{Li}},
  \bibinfo{journal}{Phys. Rev. D} \textbf{\bibinfo{volume}{22}},
  \bibinfo{pages}{2860} (\bibinfo{year}{1980}),
  \urlprefix\url{https://link.aps.org/doi/10.1103/PhysRevD.22.2860}.

\bibitem[{\citenamefont{Mohapatra and Senjanovi\ifmmode~\acute{c}\else
  \'{c}\fi{}}(1980)}]{PhysRevLett.44.912}
\bibinfo{author}{\bibfnamefont{R.~N.} \bibnamefont{Mohapatra}}
  \bibnamefont{and}
  \bibinfo{author}{\bibfnamefont{G.}~\bibnamefont{Senjanovi\ifmmode~\acute{c}\else
  \'{c}\fi{}}}, \bibinfo{journal}{Phys. Rev. Lett.}
  \textbf{\bibinfo{volume}{44}}, \bibinfo{pages}{912} (\bibinfo{year}{1980}),
  \urlprefix\url{https://link.aps.org/doi/10.1103/PhysRevLett.44.912}.

\bibitem[{\citenamefont{Ma}(1998)}]{Ma:1998dn}
\bibinfo{author}{\bibfnamefont{E.}~\bibnamefont{Ma}}, \bibinfo{journal}{Phys.
  Rev. Lett.} \textbf{\bibinfo{volume}{81}}, \bibinfo{pages}{1171}
  (\bibinfo{year}{1998}), \eprint{hep-ph/9805219}.

\bibitem[{\citenamefont{Foot et~al.}(1989)\citenamefont{Foot, Lew, He, and
  Joshi}}]{Foot:1988aq}
\bibinfo{author}{\bibfnamefont{R.}~\bibnamefont{Foot}},
  \bibinfo{author}{\bibfnamefont{H.}~\bibnamefont{Lew}},
  \bibinfo{author}{\bibfnamefont{X.~G.} \bibnamefont{He}}, \bibnamefont{and}
  \bibinfo{author}{\bibfnamefont{G.~C.} \bibnamefont{Joshi}},
  \bibinfo{journal}{Z. Phys. C} \textbf{\bibinfo{volume}{44}},
  \bibinfo{pages}{441} (\bibinfo{year}{1989}).

\bibitem[{\citenamefont{Ma and Sarkar}(1998)}]{Ma:1998dx}
\bibinfo{author}{\bibfnamefont{E.}~\bibnamefont{Ma}} \bibnamefont{and}
  \bibinfo{author}{\bibfnamefont{U.}~\bibnamefont{Sarkar}},
  \bibinfo{journal}{Phys. Rev. Lett.} \textbf{\bibinfo{volume}{80}},
  \bibinfo{pages}{5716} (\bibinfo{year}{1998}), \eprint{hep-ph/9802445}.

\bibitem[{\citenamefont{Ma and Roy}(2002)}]{Ma:2002pf}
\bibinfo{author}{\bibfnamefont{E.}~\bibnamefont{Ma}} \bibnamefont{and}
  \bibinfo{author}{\bibfnamefont{D.~P.} \bibnamefont{Roy}},
  \bibinfo{journal}{Nucl. Phys. B} \textbf{\bibinfo{volume}{644}},
  \bibinfo{pages}{290} (\bibinfo{year}{2002}), \eprint{hep-ph/0206150}.

\bibitem[{\citenamefont{Hambye et~al.}(2004)\citenamefont{Hambye, Lin, Notari,
  Papucci, and Strumia}}]{Hambye:2003rt}
\bibinfo{author}{\bibfnamefont{T.}~\bibnamefont{Hambye}},
  \bibinfo{author}{\bibfnamefont{Y.}~\bibnamefont{Lin}},
  \bibinfo{author}{\bibfnamefont{A.}~\bibnamefont{Notari}},
  \bibinfo{author}{\bibfnamefont{M.}~\bibnamefont{Papucci}}, \bibnamefont{and}
  \bibinfo{author}{\bibfnamefont{A.}~\bibnamefont{Strumia}},
  \bibinfo{journal}{Nucl. Phys. B} \textbf{\bibinfo{volume}{695}},
  \bibinfo{pages}{169} (\bibinfo{year}{2004}), \eprint{hep-ph/0312203}.

\bibitem[{\citenamefont{Okada et~al.}(2019)\citenamefont{Okada, Okada, and
  Raut}}]{Okada:2018tgy}
\bibinfo{author}{\bibfnamefont{N.}~\bibnamefont{Okada}},
  \bibinfo{author}{\bibfnamefont{S.}~\bibnamefont{Okada}}, \bibnamefont{and}
  \bibinfo{author}{\bibfnamefont{D.}~\bibnamefont{Raut}},
  \bibinfo{journal}{Phys. Rev. D} \textbf{\bibinfo{volume}{100}},
  \bibinfo{pages}{035022} (\bibinfo{year}{2019}), \eprint{1811.11927}.

\bibitem[{\citenamefont{Okada and Seto}(2020)}]{Okada:2019sbb}
\bibinfo{author}{\bibfnamefont{N.}~\bibnamefont{Okada}} \bibnamefont{and}
  \bibinfo{author}{\bibfnamefont{O.}~\bibnamefont{Seto}},
  \bibinfo{journal}{Phys. Rev. D} \textbf{\bibinfo{volume}{101}},
  \bibinfo{pages}{023522} (\bibinfo{year}{2020}), \eprint{1908.09277}.

\bibitem[{\citenamefont{Okada et~al.}(2020)\citenamefont{Okada, Okada, Raut,
  and Shafi}}]{Okada:2020evk}
\bibinfo{author}{\bibfnamefont{N.}~\bibnamefont{Okada}},
  \bibinfo{author}{\bibfnamefont{S.}~\bibnamefont{Okada}},
  \bibinfo{author}{\bibfnamefont{D.}~\bibnamefont{Raut}}, \bibnamefont{and}
  \bibinfo{author}{\bibfnamefont{Q.}~\bibnamefont{Shafi}},
  \bibinfo{journal}{Phys. Lett. B} \textbf{\bibinfo{volume}{810}},
  \bibinfo{pages}{135785} (\bibinfo{year}{2020}), \eprint{2007.02898}.

\bibitem[{\citenamefont{Ashanujjaman and Ghosh}(2022)}]{Ashanujjaman:2021txz}
\bibinfo{author}{\bibfnamefont{S.}~\bibnamefont{Ashanujjaman}}
  \bibnamefont{and} \bibinfo{author}{\bibfnamefont{K.}~\bibnamefont{Ghosh}},
  \bibinfo{journal}{JHEP} \textbf{\bibinfo{volume}{03}}, \bibinfo{pages}{195}
  (\bibinfo{year}{2022}), \eprint{2108.10952}.

\bibitem[{\citenamefont{Okada and Seto}(2022)}]{Okada:2022cby}
\bibinfo{author}{\bibfnamefont{N.}~\bibnamefont{Okada}} \bibnamefont{and}
  \bibinfo{author}{\bibfnamefont{O.}~\bibnamefont{Seto}},
  \bibinfo{journal}{Phys. Rev. D} \textbf{\bibinfo{volume}{105}},
  \bibinfo{pages}{123512} (\bibinfo{year}{2022}), \eprint{2202.08508}.

\bibitem[{\citenamefont{Ghosh et~al.}(2022)\citenamefont{Ghosh, Mahapatra,
  Narendra, and Sahu}}]{Ghosh:2021khk}
\bibinfo{author}{\bibfnamefont{P.}~\bibnamefont{Ghosh}},
  \bibinfo{author}{\bibfnamefont{S.}~\bibnamefont{Mahapatra}},
  \bibinfo{author}{\bibfnamefont{N.}~\bibnamefont{Narendra}}, \bibnamefont{and}
  \bibinfo{author}{\bibfnamefont{N.}~\bibnamefont{Sahu}},
  \bibinfo{journal}{Phys. Rev. D} \textbf{\bibinfo{volume}{106}},
  \bibinfo{pages}{015001} (\bibinfo{year}{2022}), \eprint{2107.11951}.

\bibitem[{\citenamefont{Aad et~al.}(2024)}]{ATLAS:2023yqk}
\bibinfo{author}{\bibfnamefont{G.}~\bibnamefont{Aad}} \bibnamefont{et~al.}
  (\bibinfo{collaboration}{ATLAS, CMS}), \bibinfo{journal}{Phys. Rev. Lett.}
  \textbf{\bibinfo{volume}{132}}, \bibinfo{pages}{021803}
  (\bibinfo{year}{2024}), \eprint{2309.03501}.

\bibitem[{\citenamefont{Barducci et~al.}(2023)\citenamefont{Barducci, Di~Luzio,
  Nardecchia, and Toni}}]{Barducci:2023zml}
\bibinfo{author}{\bibfnamefont{D.}~\bibnamefont{Barducci}},
  \bibinfo{author}{\bibfnamefont{L.}~\bibnamefont{Di~Luzio}},
  \bibinfo{author}{\bibfnamefont{M.}~\bibnamefont{Nardecchia}},
  \bibnamefont{and} \bibinfo{author}{\bibfnamefont{C.}~\bibnamefont{Toni}},
  \bibinfo{journal}{JHEP} \textbf{\bibinfo{volume}{12}}, \bibinfo{pages}{154}
  (\bibinfo{year}{2023}), \eprint{2311.10130}.

\bibitem[{\citenamefont{Boto et~al.}(2023)\citenamefont{Boto, Das, Romao, Saha,
  and Silva}}]{Boto:2023bpg}
\bibinfo{author}{\bibfnamefont{R.}~\bibnamefont{Boto}},
  \bibinfo{author}{\bibfnamefont{D.}~\bibnamefont{Das}},
  \bibinfo{author}{\bibfnamefont{J.~C.} \bibnamefont{Romao}},
  \bibinfo{author}{\bibfnamefont{I.}~\bibnamefont{Saha}}, \bibnamefont{and}
  \bibinfo{author}{\bibfnamefont{J.~P.} \bibnamefont{Silva}}
  (\bibinfo{year}{2023}), \eprint{2312.13050}.

\bibitem[{\citenamefont{Arbabifar et~al.}(2013)\citenamefont{Arbabifar,
  Bahrami, and Frank}}]{Arbabifar:2012bd}
\bibinfo{author}{\bibfnamefont{F.}~\bibnamefont{Arbabifar}},
  \bibinfo{author}{\bibfnamefont{S.}~\bibnamefont{Bahrami}}, \bibnamefont{and}
  \bibinfo{author}{\bibfnamefont{M.}~\bibnamefont{Frank}},
  \bibinfo{journal}{Phys. Rev. D} \textbf{\bibinfo{volume}{87}},
  \bibinfo{pages}{015020} (\bibinfo{year}{2013}), \eprint{1211.6797}.

\bibitem[{\citenamefont{Bhupal~Dev et~al.}(2013)\citenamefont{Bhupal~Dev,
  Ghosh, Okada, and Saha}}]{BhupalDev:2013xol}
\bibinfo{author}{\bibfnamefont{P.~S.} \bibnamefont{Bhupal~Dev}},
  \bibinfo{author}{\bibfnamefont{D.~K.} \bibnamefont{Ghosh}},
  \bibinfo{author}{\bibfnamefont{N.}~\bibnamefont{Okada}}, \bibnamefont{and}
  \bibinfo{author}{\bibfnamefont{I.}~\bibnamefont{Saha}},
  \bibinfo{journal}{JHEP} \textbf{\bibinfo{volume}{03}}, \bibinfo{pages}{150}
  (\bibinfo{year}{2013}), \bibinfo{note}{[Erratum: JHEP 05, 049 (2013)]},
  \eprint{1301.3453}.

\bibitem[{\citenamefont{Ma}(2017)}]{Ma:2017ucp}
\bibinfo{author}{\bibfnamefont{E.}~\bibnamefont{Ma}}, \bibinfo{journal}{Phys.
  Lett. B} \textbf{\bibinfo{volume}{772}}, \bibinfo{pages}{442}
  (\bibinfo{year}{2017}), \eprint{1704.04666}.

\bibitem[{LEP(2001)}]{LEPHiggsWorkingGroupforHiggsbosonsearches:2001ogs}
in \emph{\bibinfo{booktitle}{{2001 Europhysics Conference on High Energy
  Physics}}} (\bibinfo{year}{2001}), \eprint{hep-ex/0107031}.

\bibitem[{\citenamefont{Aad et~al.}(2023{\natexlab{a}})}]{ATLAS:2022pbd}
\bibinfo{author}{\bibfnamefont{G.}~\bibnamefont{Aad}} \bibnamefont{et~al.}
  (\bibinfo{collaboration}{ATLAS}), \bibinfo{journal}{Eur. Phys. J. C}
  \textbf{\bibinfo{volume}{83}}, \bibinfo{pages}{605}
  (\bibinfo{year}{2023}{\natexlab{a}}), \eprint{2211.07505}.

\bibitem[{\citenamefont{Bhattacharya et~al.}(2017)\citenamefont{Bhattacharya,
  Sahoo, and Sahu}}]{Bhattacharya:2017sml}
\bibinfo{author}{\bibfnamefont{S.}~\bibnamefont{Bhattacharya}},
  \bibinfo{author}{\bibfnamefont{N.}~\bibnamefont{Sahoo}}, \bibnamefont{and}
  \bibinfo{author}{\bibfnamefont{N.}~\bibnamefont{Sahu}},
  \bibinfo{journal}{Phys. Rev. D} \textbf{\bibinfo{volume}{96}},
  \bibinfo{pages}{035010} (\bibinfo{year}{2017}), \eprint{1704.03417}.

\bibitem[{\citenamefont{L\'opez-Val and Robens}(2014)}]{Lopez-Val:2014jva}
\bibinfo{author}{\bibfnamefont{D.}~\bibnamefont{L\'opez-Val}} \bibnamefont{and}
  \bibinfo{author}{\bibfnamefont{T.}~\bibnamefont{Robens}},
  \bibinfo{journal}{Phys. Rev. D} \textbf{\bibinfo{volume}{90}},
  \bibinfo{pages}{114018} (\bibinfo{year}{2014}), \eprint{1406.1043}.

\bibitem[{\citenamefont{Robens and Stefaniak}(2016)}]{Robens:2016xkb}
\bibinfo{author}{\bibfnamefont{T.}~\bibnamefont{Robens}} \bibnamefont{and}
  \bibinfo{author}{\bibfnamefont{T.}~\bibnamefont{Stefaniak}},
  \bibinfo{journal}{Eur. Phys. J. C} \textbf{\bibinfo{volume}{76}},
  \bibinfo{pages}{268} (\bibinfo{year}{2016}), \eprint{1601.07880}.

\bibitem[{\citenamefont{Bellgardt et~al.}(1988)}]{SINDRUM:1987nra}
\bibinfo{author}{\bibfnamefont{U.}~\bibnamefont{Bellgardt}}
  \bibnamefont{et~al.} (\bibinfo{collaboration}{SINDRUM}),
  \bibinfo{journal}{Nucl. Phys. B} \textbf{\bibinfo{volume}{299}},
  \bibinfo{pages}{1} (\bibinfo{year}{1988}).

\bibitem[{\citenamefont{Baldini et~al.}(2016)}]{MEG:2016leq}
\bibinfo{author}{\bibfnamefont{A.~M.} \bibnamefont{Baldini}}
  \bibnamefont{et~al.} (\bibinfo{collaboration}{MEG}), \bibinfo{journal}{Eur.
  Phys. J. C} \textbf{\bibinfo{volume}{76}}, \bibinfo{pages}{434}
  (\bibinfo{year}{2016}), \eprint{1605.05081}.

\bibitem[{\citenamefont{Chun et~al.}(2012)\citenamefont{Chun, Lee, and
  Sharma}}]{Chun:2012jw}
\bibinfo{author}{\bibfnamefont{E.~J.} \bibnamefont{Chun}},
  \bibinfo{author}{\bibfnamefont{H.~M.} \bibnamefont{Lee}}, \bibnamefont{and}
  \bibinfo{author}{\bibfnamefont{P.}~\bibnamefont{Sharma}},
  \bibinfo{journal}{JHEP} \textbf{\bibinfo{volume}{11}}, \bibinfo{pages}{106}
  (\bibinfo{year}{2012}), \eprint{1209.1303}.

\bibitem[{\citenamefont{Aad et~al.}(2022)}]{ATLAS:2022yvh}
\bibinfo{author}{\bibfnamefont{G.}~\bibnamefont{Aad}} \bibnamefont{et~al.}
  (\bibinfo{collaboration}{ATLAS}), \bibinfo{journal}{JHEP}
  \textbf{\bibinfo{volume}{08}}, \bibinfo{pages}{104} (\bibinfo{year}{2022}),
  \eprint{2202.07953}.

\bibitem[{\citenamefont{Shifman et~al.}(1979)\citenamefont{Shifman, Vainshtein,
  Voloshin, and Zakharov}}]{Shifman:1979eb}
\bibinfo{author}{\bibfnamefont{M.~A.} \bibnamefont{Shifman}},
  \bibinfo{author}{\bibfnamefont{A.~I.} \bibnamefont{Vainshtein}},
  \bibinfo{author}{\bibfnamefont{M.~B.} \bibnamefont{Voloshin}},
  \bibnamefont{and} \bibinfo{author}{\bibfnamefont{V.~I.}
  \bibnamefont{Zakharov}}, \bibinfo{journal}{Sov. J. Nucl. Phys.}
  \textbf{\bibinfo{volume}{30}}, \bibinfo{pages}{711} (\bibinfo{year}{1979}).

\bibitem[{\citenamefont{Djouadi}(2008)}]{Djouadi:2005gj}
\bibinfo{author}{\bibfnamefont{A.}~\bibnamefont{Djouadi}},
  \bibinfo{journal}{Phys. Rept.} \textbf{\bibinfo{volume}{459}},
  \bibinfo{pages}{1} (\bibinfo{year}{2008}), \eprint{hep-ph/0503173}.

\bibitem[{\citenamefont{Spira}(1998)}]{Spira:1997dg}
\bibinfo{author}{\bibfnamefont{M.}~\bibnamefont{Spira}},
  \bibinfo{journal}{Fortsch. Phys.} \textbf{\bibinfo{volume}{46}},
  \bibinfo{pages}{203} (\bibinfo{year}{1998}), \eprint{hep-ph/9705337}.

\bibitem[{\citenamefont{Gunion et~al.}(2000)\citenamefont{Gunion, Haber, Kane,
  and Dawson}}]{Gunion:1989we}
\bibinfo{author}{\bibfnamefont{J.~F.} \bibnamefont{Gunion}},
  \bibinfo{author}{\bibfnamefont{H.~E.} \bibnamefont{Haber}},
  \bibinfo{author}{\bibfnamefont{G.~L.} \bibnamefont{Kane}}, \bibnamefont{and}
  \bibinfo{author}{\bibfnamefont{S.}~\bibnamefont{Dawson}},
  \emph{\bibinfo{title}{{The Higgs Hunter's Guide}}}, vol.~\bibinfo{volume}{80}
  (\bibinfo{year}{2000}), ISBN \bibinfo{isbn}{978-0-429-49644-8}.

\bibitem[{\citenamefont{Carena et~al.}(2012)\citenamefont{Carena, Low, and
  Wagner}}]{Carena:2012xa}
\bibinfo{author}{\bibfnamefont{M.}~\bibnamefont{Carena}},
  \bibinfo{author}{\bibfnamefont{I.}~\bibnamefont{Low}}, \bibnamefont{and}
  \bibinfo{author}{\bibfnamefont{C.~E.~M.} \bibnamefont{Wagner}},
  \bibinfo{journal}{JHEP} \textbf{\bibinfo{volume}{08}}, \bibinfo{pages}{060}
  (\bibinfo{year}{2012}), \eprint{1206.1082}.

\bibitem[{\citenamefont{Aad et~al.}(2023{\natexlab{b}})}]{ATLAS:2022tnm}
\bibinfo{author}{\bibfnamefont{G.}~\bibnamefont{Aad}} \bibnamefont{et~al.}
  (\bibinfo{collaboration}{ATLAS}), \bibinfo{journal}{JHEP}
  \textbf{\bibinfo{volume}{07}}, \bibinfo{pages}{088}
  (\bibinfo{year}{2023}{\natexlab{b}}), \eprint{2207.00348}.

\bibitem[{\citenamefont{Aad et~al.}(2020)}]{ATLAS:2020qcv}
\bibinfo{author}{\bibfnamefont{G.}~\bibnamefont{Aad}} \bibnamefont{et~al.}
  (\bibinfo{collaboration}{ATLAS}), \bibinfo{journal}{Phys. Lett. B}
  \textbf{\bibinfo{volume}{809}}, \bibinfo{pages}{135754}
  (\bibinfo{year}{2020}), \eprint{2005.05382}.

\bibitem[{\citenamefont{Sirunyan et~al.}(2021)}]{CMS:2021kom}
\bibinfo{author}{\bibfnamefont{A.~M.} \bibnamefont{Sirunyan}}
  \bibnamefont{et~al.} (\bibinfo{collaboration}{CMS}), \bibinfo{journal}{JHEP}
  \textbf{\bibinfo{volume}{07}}, \bibinfo{pages}{027} (\bibinfo{year}{2021}),
  \eprint{2103.06956}.

\bibitem[{\citenamefont{Tumasyan et~al.}(2023)}]{CMS:2022ahq}
\bibinfo{author}{\bibfnamefont{A.}~\bibnamefont{Tumasyan}} \bibnamefont{et~al.}
  (\bibinfo{collaboration}{CMS}), \bibinfo{journal}{JHEP}
  \textbf{\bibinfo{volume}{05}}, \bibinfo{pages}{233} (\bibinfo{year}{2023}),
  \eprint{2204.12945}.

\bibitem[{\citenamefont{Lee et~al.}(1977)\citenamefont{Lee, Quigg, and
  Thacker}}]{PhysRevD.16.1519}
\bibinfo{author}{\bibfnamefont{B.~W.} \bibnamefont{Lee}},
  \bibinfo{author}{\bibfnamefont{C.}~\bibnamefont{Quigg}}, \bibnamefont{and}
  \bibinfo{author}{\bibfnamefont{H.~B.} \bibnamefont{Thacker}},
  \bibinfo{journal}{Phys. Rev. D} \textbf{\bibinfo{volume}{16}},
  \bibinfo{pages}{1519} (\bibinfo{year}{1977}),
  \urlprefix\url{https://link.aps.org/doi/10.1103/PhysRevD.16.1519}.

\bibitem[{\citenamefont{Aoki and Kanemura}(2008)}]{PhysRevD.77.095009}
\bibinfo{author}{\bibfnamefont{M.}~\bibnamefont{Aoki}} \bibnamefont{and}
  \bibinfo{author}{\bibfnamefont{S.}~\bibnamefont{Kanemura}},
  \bibinfo{journal}{Phys. Rev. D} \textbf{\bibinfo{volume}{77}},
  \bibinfo{pages}{095009} (\bibinfo{year}{2008}),
  \urlprefix\url{https://link.aps.org/doi/10.1103/PhysRevD.77.095009}.

\bibitem[{\citenamefont{{G Akeroyd} et~al.}(2000)\citenamefont{{G Akeroyd},
  Arhrib, and Naimi}}]{GAKEROYD2000119}
\bibinfo{author}{\bibfnamefont{A.}~\bibnamefont{{G Akeroyd}}},
  \bibinfo{author}{\bibfnamefont{A.}~\bibnamefont{Arhrib}}, \bibnamefont{and}
  \bibinfo{author}{\bibfnamefont{E.}~\bibnamefont{Naimi}},
  \bibinfo{journal}{Physics Letters B} \textbf{\bibinfo{volume}{490}},
  \bibinfo{pages}{119} (\bibinfo{year}{2000}), ISSN \bibinfo{issn}{0370-2693},
  \urlprefix\url{https://www.sciencedirect.com/science/article/pii/S037026930000962X}.

\bibitem[{\citenamefont{Ghosh et~al.}(2018)\citenamefont{Ghosh, Ghosh, Saha,
  and Shaw}}]{Ghosh:2017pxl}
\bibinfo{author}{\bibfnamefont{D.~K.} \bibnamefont{Ghosh}},
  \bibinfo{author}{\bibfnamefont{N.}~\bibnamefont{Ghosh}},
  \bibinfo{author}{\bibfnamefont{I.}~\bibnamefont{Saha}}, \bibnamefont{and}
  \bibinfo{author}{\bibfnamefont{A.}~\bibnamefont{Shaw}},
  \bibinfo{journal}{Phys. Rev. D} \textbf{\bibinfo{volume}{97}},
  \bibinfo{pages}{115022} (\bibinfo{year}{2018}), \eprint{1711.06062}.

\bibitem[{\citenamefont{Belanger et~al.}(2007)\citenamefont{Belanger, Boudjema,
  Pukhov, and Semenov}}]{Belanger:2006is}
\bibinfo{author}{\bibfnamefont{G.}~\bibnamefont{Belanger}},
  \bibinfo{author}{\bibfnamefont{F.}~\bibnamefont{Boudjema}},
  \bibinfo{author}{\bibfnamefont{A.}~\bibnamefont{Pukhov}}, \bibnamefont{and}
  \bibinfo{author}{\bibfnamefont{A.}~\bibnamefont{Semenov}},
  \bibinfo{journal}{Comput. Phys. Commun.} \textbf{\bibinfo{volume}{176}},
  \bibinfo{pages}{367} (\bibinfo{year}{2007}), \eprint{hep-ph/0607059}.

\bibitem[{\citenamefont{Alloul et~al.}(2014)\citenamefont{Alloul, Christensen,
  Degrande, Duhr, and Fuks}}]{Alloul:2013bka}
\bibinfo{author}{\bibfnamefont{A.}~\bibnamefont{Alloul}},
  \bibinfo{author}{\bibfnamefont{N.~D.} \bibnamefont{Christensen}},
  \bibinfo{author}{\bibfnamefont{C.}~\bibnamefont{Degrande}},
  \bibinfo{author}{\bibfnamefont{C.}~\bibnamefont{Duhr}}, \bibnamefont{and}
  \bibinfo{author}{\bibfnamefont{B.}~\bibnamefont{Fuks}},
  \bibinfo{journal}{Comput. Phys. Commun.} \textbf{\bibinfo{volume}{185}},
  \bibinfo{pages}{2250} (\bibinfo{year}{2014}), \eprint{1310.1921}.

\bibitem[{\citenamefont{Belanger et~al.}(2009)\citenamefont{Belanger, Boudjema,
  Pukhov, and Semenov}}]{Belanger:2008sj}
\bibinfo{author}{\bibfnamefont{G.}~\bibnamefont{Belanger}},
  \bibinfo{author}{\bibfnamefont{F.}~\bibnamefont{Boudjema}},
  \bibinfo{author}{\bibfnamefont{A.}~\bibnamefont{Pukhov}}, \bibnamefont{and}
  \bibinfo{author}{\bibfnamefont{A.}~\bibnamefont{Semenov}},
  \bibinfo{journal}{Comput. Phys. Commun.} \textbf{\bibinfo{volume}{180}},
  \bibinfo{pages}{747} (\bibinfo{year}{2009}), \eprint{0803.2360}.

\bibitem[{\citenamefont{Goodenough and Hooper}(2009)}]{Goodenough:2009gk}
\bibinfo{author}{\bibfnamefont{L.}~\bibnamefont{Goodenough}} \bibnamefont{and}
  \bibinfo{author}{\bibfnamefont{D.}~\bibnamefont{Hooper}}
  (\bibinfo{year}{2009}), \eprint{0910.2998}.

\bibitem[{\citenamefont{Elor et~al.}(2016)\citenamefont{Elor, Rodd, Slatyer,
  and Xue}}]{Elor:2015bho}
\bibinfo{author}{\bibfnamefont{G.}~\bibnamefont{Elor}},
  \bibinfo{author}{\bibfnamefont{N.~L.} \bibnamefont{Rodd}},
  \bibinfo{author}{\bibfnamefont{T.~R.} \bibnamefont{Slatyer}},
  \bibnamefont{and} \bibinfo{author}{\bibfnamefont{W.}~\bibnamefont{Xue}},
  \bibinfo{journal}{JCAP} \textbf{\bibinfo{volume}{06}}, \bibinfo{pages}{024}
  (\bibinfo{year}{2016}), \eprint{1511.08787}.

\bibitem[{\citenamefont{Babu et~al.}(2023)\citenamefont{Babu, Chakdar, Das,
  Ghosh, and Ghosh}}]{Babu:2023zni}
\bibinfo{author}{\bibfnamefont{K.~S.} \bibnamefont{Babu}},
  \bibinfo{author}{\bibfnamefont{S.}~\bibnamefont{Chakdar}},
  \bibinfo{author}{\bibfnamefont{N.}~\bibnamefont{Das}},
  \bibinfo{author}{\bibfnamefont{D.~K.} \bibnamefont{Ghosh}}, \bibnamefont{and}
  \bibinfo{author}{\bibfnamefont{P.}~\bibnamefont{Ghosh}},
  \bibinfo{journal}{JHEP} \textbf{\bibinfo{volume}{07}}, \bibinfo{pages}{143}
  (\bibinfo{year}{2023}), \eprint{2305.03167}.

\bibitem[{\citenamefont{Arhrib et~al.}(2012)\citenamefont{Arhrib, Benbrik,
  Chabab, Moultaka, and Rahili}}]{Arhrib:2011vc}
\bibinfo{author}{\bibfnamefont{A.}~\bibnamefont{Arhrib}},
  \bibinfo{author}{\bibfnamefont{R.}~\bibnamefont{Benbrik}},
  \bibinfo{author}{\bibfnamefont{M.}~\bibnamefont{Chabab}},
  \bibinfo{author}{\bibfnamefont{G.}~\bibnamefont{Moultaka}}, \bibnamefont{and}
  \bibinfo{author}{\bibfnamefont{L.}~\bibnamefont{Rahili}},
  \bibinfo{journal}{JHEP} \textbf{\bibinfo{volume}{04}}, \bibinfo{pages}{136}
  (\bibinfo{year}{2012}), \eprint{1112.5453}.

\bibitem[{\citenamefont{Arhrib et~al.}(2011)\citenamefont{Arhrib, Benbrik,
  Chabab, Moultaka, Peyranere, Rahili, and Ramadan}}]{Arhrib:2011uy}
\bibinfo{author}{\bibfnamefont{A.}~\bibnamefont{Arhrib}},
  \bibinfo{author}{\bibfnamefont{R.}~\bibnamefont{Benbrik}},
  \bibinfo{author}{\bibfnamefont{M.}~\bibnamefont{Chabab}},
  \bibinfo{author}{\bibfnamefont{G.}~\bibnamefont{Moultaka}},
  \bibinfo{author}{\bibfnamefont{M.~C.} \bibnamefont{Peyranere}},
  \bibinfo{author}{\bibfnamefont{L.}~\bibnamefont{Rahili}}, \bibnamefont{and}
  \bibinfo{author}{\bibfnamefont{J.}~\bibnamefont{Ramadan}},
  \bibinfo{journal}{Phys. Rev. D} \textbf{\bibinfo{volume}{84}},
  \bibinfo{pages}{095005} (\bibinfo{year}{2011}), \eprint{1105.1925}.

\end{thebibliography}
\end{document}